\documentclass{sig-alternate}

\usepackage[german,american]{babel}
\usepackage[T1]{fontenc}
\usepackage[utf8]{inputenc}
\usepackage{times}
\usepackage{microtype}

\makeatletter
\def\@copyrightspace{\relax}

\long\def\@makecaption#1#2{
   \vskip 4pt 
   \setbox\@tempboxa\hbox{\textbf{#1: #2}}
   \ifdim \wd\@tempboxa >\hsize 
       \textbf{#1: #2}\par                
     \else                      
       \hbox to\hsize{\hfil\box\@tempboxa\hfil}
   \fi}
\makeatother

\renewcommand{\tt}{\fontencoding{OT1}\fontfamily{cmtt}\selectfont}
\usepackage{pifont} 

\newcommand{\bslabel}[1]{\textbf{#1.}}

\newcommand{\Ni}{(1)~}
\newcommand{\Nii}{(2)~}
\newcommand{\Niii}{(3)~}
\newcommand{\Niv}{(4)~}
\newcommand{\Nv}{(5)~}

\usepackage{graphicx}
\DeclareGraphicsExtensions{.pdf,.ai,.jpg,.png}
\setkeys{Gin}{pagebox=artbox}
\graphicspath{{figures/}}

\newcommand{\bsfigure}[3][scale=1.0]{%
  \begin{figure}[tb]
    \centering
    \includegraphics[#1]{#2}
    \caption{#3}\label{#2}
  \end{figure}}

\usepackage{multirow}
\usepackage{arydshln}
\usepackage{booktabs}
\usepackage[numbers,sectionbib]{natbib}
\usepackage[usenames,dvipsnames]{color}
\RequirePackage{soul}
\setstcolor{blue}

\usepackage{xspace}
\newcommand{\rocauc}{ROC\ensuremath{_\textrm{AUC}}\xspace}
\newcommand{\prauc}{PR\ensuremath{_\textrm{AUC}}\xspace}

\newcommand{\lowerCaseRatio}{{\tt low\-er\-Case\-Ra\-tio}\xspace}
\newcommand{\upperCaseRatio}{{\tt up\-per\-Case\-Ra\-tio}\xspace}
\newcommand{\latinRatio}{{\tt la\-tin\-Ra\-tio}\xspace}
\newcommand{\nonLatinRatio}{{\tt non\-La\-tin\-Ratio}\xspace}
\newcommand{\alphanumericRatio}{{\tt al\-pha\-nu\-me\-ric\-Ra\-tio}\xspace}
\newcommand{\digitRatio}{{\tt di\-git\-Ra\-tio}\xspace}
\newcommand{\punctuationRatio}{{\tt punc\-tu\-a\-tion\-Ra\-tio}\xspace}
\newcommand{\whitespaceRatio}{{\tt white\-space\-Ra\-tio}\xspace}
\newcommand{\longestCharacterSequence}{{\tt long\-est\-Cha\-rac\-ter\-Seq\-uence}\xspace}
\newcommand{\asciiRatio}{{\tt ascii\-Ra\-tio}\xspace}
\newcommand{\bracketRatio}{{\tt brac\-ket\-Ra\-tio}\xspace}
\newcommand{\symbolRatio}{{\tt sym\-bol\-Ra\-tio}\xspace}
\newcommand{\mainAlphabet}{{\tt main\-Al\-pha\-bet}\xspace}

\newcommand{\languageWordRatio}{{\tt lang\-uage\-Word\-Ra\-tio}\xspace}
\newcommand{\containsLanguageWord}{{\tt con\-tains\-Lang\-uage\-Word}\xspace}
\newcommand{\lowerCaseWordRatio}{{\tt low\-er\-Case\-Word\-Ra\-tio}\xspace}
\newcommand{\longestWord}{{\tt long\-est\-Word}\xspace}
\newcommand{\containsURL}{{\tt con\-tains\-URL}\xspace}
\newcommand{\badWordRatio}{{\tt bad\-Word\-Ra\-tio}\xspace}
\newcommand{\proportionOfQidAdded}{{\tt pro\-por\-tion\-Of\-Qid\-Ad\-ded}\xspace}
\newcommand{\upperCaseWordRatio}{{\tt up\-per\-Case\-Word\-Ra\-tio}\xspace}
\newcommand{\proportionOfLinksAdded}{{\tt pro\-por\-tion\-Of\-Links\-Ad\-ded}\xspace}
\newcommand{\bagOfWords}{{\tt bag\-Of\-Words}\xspace}
\newcommand{\proportionOfLanguageAdded}{{\tt pro\-por\-tion\-Of\-Lang\-uage\-Ad\-ded}\xspace}

\newcommand{\commentTailLength}{{\tt com\-ment\-Tail\-Length}\xspace}
\newcommand{\commentSitelinkSimilarity}{{\tt com\-ment\-Site\-link\-Si\-mi\-la\-ri\-ty}\xspace}
\newcommand{\commentLabelSimilarity}{{\tt com\-ment\-La\-bel\-Si\-mi\-la\-ri\-ty}\xspace}
\newcommand{\commentCommentSimilarity}{{\tt com\-ment\-Com\-ment\-Si\-mi\-la\-ri\-ty}\xspace}
\newcommand{\languageMatchProb}{{\tt lang\-uage\-Match\-Prob}\xspace}
\newcommand{\hasIdentifierChanged}{{\tt has\-Identifier\-Chan\-ged}\xspace}

\newcommand{\propertyFrequency}{{\tt pro\-per\-ty\-Fre\-quen\-cy}\xspace}
\newcommand{\itemValueFrequency}{{\tt item\-Val\-ue\-Fre\-quen\-cy}\xspace}
\newcommand{\literalValueFrequency}{{\tt lit\-e\-ral\-Val\-ue\-Fre\-quen\-cy}\xspace}

\newcommand{\userCountry}{{\tt user\-Coun\-try}\xspace}
\newcommand{\userTimeZone}{{\tt user\-Time\-Zone}\xspace}
\newcommand{\userCity}{{\tt user\-City}\xspace}
\newcommand{\userCounty}{{\tt user\-County}\xspace}
\newcommand{\userRegion}{{\tt user\-Re\-gion}\xspace}
\newcommand{\cumUserUniqueItems}{{\tt cum\-User\-Unique\-Items}\xspace}
\newcommand{\userContinent}{{\tt user\-Con\-ti\-nent}\xspace}
\newcommand{\isRegisteredUser}{{\tt is\-Re\-gis\-tered\-User}\xspace}
\newcommand{\userFrequency}{{\tt user\-Fre\-quen\-cy}\xspace}
\newcommand{\isPrivilegedUser}{{\tt is\-Pri\-vi\-leged\-User}\xspace}
\newcommand{\userIPSubnets}{{\tt user\-IP\-Sub\-nets}\xspace}
\newcommand{\userVandalismFraction}{{\tt user\-Van\-\-da\-lism\-Frac\-tion}\xspace}
\newcommand{\userVandalismCount}{{\tt user\-Van\-\-da\-lism\-Count}\xspace}
\newcommand{\userUniqueItems}{{\tt user\-Unique\-Items}\xspace}
\newcommand{\userAge}{{\tt user\-Age}\xspace}

\newcommand{\logCumItemUniqueUsers}{{\tt log\-Cum\-Item\-Unique\-Users}\xspace}
\newcommand{\logItemFrequency}{{\tt log\-Item\-Fre\-quen\-cy}\xspace}

\newcommand{\isHuman}{{\tt is\-Hu\-man}\xspace}

\newcommand{\itemFrequency}{{\tt item\-Fre\-quen\-cy}\xspace}
\newcommand{\itemVandalismFraction}{{\tt item\-Van\-\-da\-lism\-Frac\-tion}\xspace}
\newcommand{\itemVandalismCount}{{\tt item\-Van\-\-da\-lism\-Count}\xspace}
\newcommand{\itemUniqueUsers}{{\tt item\-Unique\-Users}\xspace}
\newcommand{\isLivingPerson}{{\tt is\-Living\-Per\-son}\xspace}

\newcommand{\revisionTags}{{\tt re\-vi\-sion\-Tags}\xspace}
\newcommand{\revisionLanguage}{{\tt re\-vi\-sion\-Lang\-uage}\xspace}
\newcommand{\revisionAction}{{\tt re\-vi\-sionAc\-tion}\xspace}
\newcommand{\commentLength}{{\tt com\-ment\-Length}\xspace}
\newcommand{\isLatinLanguage}{{\tt is\-La\-tin\-Lang\-uage}\xspace}
\newcommand{\revisionPrevAction}{{\tt re\-vi\-sion\-Prev\-Ac\-tion}\xspace}
\newcommand{\revisionSubaction}{{\tt re\-vi\-sion\-Sub\-ac\-tion}\xspace}
\newcommand{\positionWithinSession}{{\tt po\-si\-tion\-With\-in\-Ses\-sion}\xspace}
\newcommand{\numberOfIdentifiersChanged}{{\tt num\-ber\-Of\-Identifiers\-Chan\-ged}\xspace}

\newcommand{\hashTag}{{\tt hash\-Tag}\xspace}

\newcommand{\isSpecialRevision}{{\tt is\-Spe\-cial\-Re\-vi\-sion}\xspace}
\newcommand{\changeCount}{{\tt change\-Count}\xspace}
\newcommand{\isMinorRevision}{{\tt is\-Mi\-nor\-Re\-vi\-sion}\xspace}
\newcommand{\revisionSize}{{\tt re\-vi\-sion\-Size}\xspace}
\newcommand{\superItem}{{\tt su\-per\-Item}\xspace}
\newcommand{\hourOfDay}{{\tt hour\-Of\-Day}\xspace}
\newcommand{\dayOfWeek}{{\tt day\-Of\-Week}\xspace}
\newcommand{\revisionPrevUser}{{\tt re\-vi\-sion\-Prev\-User}\xspace}

\begin{document}

\title{Overview of the Wikidata Vandalism Detection Task \\[0.5ex] at WSDM Cup 2017}%

\numberofauthors{1}
\author{\alignauthor {
Stefan Heindorf$^1$ \hfill Martin Potthast$^2$ \hfill Gregor Engels$^1$ \hfill Benno Stein$^3$\\[2.5ex]%
\begin{minipage}[t]{0.33\linewidth}
\centering
\affaddr{$^1$Paderborn University}\\
\affaddr{\small <last name>@uni-paderborn.de}
\end{minipage}%
\begin{minipage}[t]{0.33\linewidth}
\centering
\affaddr{$^2$Leipzig University}\\
\affaddr{\small <first name>.<last name>@uni-leipzig.de}
\end{minipage}%
\begin{minipage}[t]{0.33\linewidth}
\centering
\affaddr{$^3$Bauhaus-Universit{\"a}t Weimar}\\
\affaddr{\small <first name>.<last name>@uni-weimar.de}
\end{minipage}}}

\maketitle

\begin{abstract}
We report on the Wikidata vandalism detection task at the WSDM Cup~2017. The task received five submissions for which this paper describes their evaluation and a comparison to state of the art baselines. Unlike previous work, we recast Wikidata vandalism detection as an online learning problem, requiring participant software to predict vandalism in near real-time. The best-performing approach achieves a \rocauc of 0.947 at a \prauc of 0.458. In particular, this task was organized as a {\em software submission task}: to maximize reproducibility as well as to foster future research and development on this task, the participants were asked to submit their working software to the TIRA experimentation platform along with the source code for open source release.
\end{abstract}

\enlargethispage{1\baselineskip}
\section{Introduction}

Knowledge is increasingly gathered by the crowd. One of the most prominent examples in this regard is Wikidata, the knowledge base of the Wikimedia Foundation. Wikidata stores knowledge (better: facts) in structured form as subject-predicate-object statements that can be edited by anyone. Most of the volunteers' contributions to Wikidata are of high quality; however, there are, just like in Wikipedia, some ``editors'' who vandalize and damage the knowledge base. The impact of these few can be severe: since Wikidata is, to an increasing extent, integrated into information systems such as search engines and question-answering systems, the risk of spreading false information to all their users increases as well. It is obvious that this threat cannot be countered by human inspection alone: currently, Wikidata gets millions of contributions every month; the effort of reviewing them manually will exceed the resources of the community, especially as Wikidata further grows.

Encouraged by the success of algorithmic vandalism detection on Wikipedia, we started a comparable endeavor for Wikidata two years ago: we carefully compiled a corpus based on Wikidata's revision history~\citep{Heindorf2015} and went on by developing the first machine learning-based Wikidata vandalism detector~\citep{Heindorf2016}. Compared with the quality of the best vandalism detectors for Wikipedia, our results may be considered as a first step toward a practical solution.

We are working on new detection approaches ourselves but we see that progress can be made at a much faster pace if independent researchers work in parallel, generating a diversity of ideas. While research communities often form around problems of interest, this has not been the case for vandalism detection in knowledge bases---perhaps due to the novelty of the task. We hence took a proactive stance by organizing a shared task event as part of the WSDM Cup~2017~\citep{Heindorf2017_WSDM}. Shared tasks have proven successful as catalysts for forming communities on a number of occasions before, in particular for vandalism detection on Wikipedia: on the basis of two shared tasks, considerable interest from researchers worldwide resulted in dozens of approaches to date~\citep{Potthast2011_Overview,Potthast2010_Overview}.

\smallskip
The goal of our shared task at the WSDM Cup~2017 is to develop an effective vandalism detection model for Wikidata:

\begin{quotation}\noindent
Given a Wikidata revision, the task is to compute a quality score denoting the likelihood of this revision being vandalism~(or, similarly, damaging).
\end{quotation}

The revisions had to be scored in near real-time as soon as they arrived, allowing for immediate action upon potential vandalism. Moreover, a model (detector) should hint at vandalism across a wide range of precision-recall-points to enable use cases such as
\Ni
fully automatic reversion of damaging edits at high precision, as well as
\Nii
pre-filtering and ranking of revisions with respect to importance of being reviewed at high recall. As our main evaluation metric we employ the area under curve of the receiver operating characteristic.

\smallskip
Our contributions to the WSDM Cup~2017 are as follows:
\vspace{-1ex}
\begin{itemize}
\item
Compilation of the Wikidata Vandalism Corpus 2016, an updated version of our previous corpus, enlarged and retrofitted for the setup of the shared task.
\item
Survey of the participant approaches with regard to features and model variants.
\item
Comparison of the participant approaches to the state of the art under a number of settings beyond the main task.
\item
Analysis of the combined performance of all models (detectors) as an ensemble in order to estimate the achievable performance when integrating all approaches.
\item
Release of an open source repository of the entire evaluation framework of the shared task, as well as release of most of the participants' code bases by courtesy of the participants.
\end{itemize}

In what follows, Section~\ref{related-work} briefly reviews the aforementioned related work, Section~\ref{evaluation-framework} introduces the developed evaluation framework including the new edition of the Wikidata vandalism corpus, Section~\ref{survey} surveys the submitted approaches, and Section~\ref{evaluation} reports on their evaluation. Finally, Section~\ref{discussion} reflects on the lessons learned.

\section{Related Work}
\label{related-work}

The section gives a comprehensive overview of the literature on vandalism detection. The two subsections detail the approaches regarding dataset construction and detection technology respectively.

\subsection{Corpus Construction}

There are basically three strategies to construct vandalism corpora for Wiki-style projects~\citep{Kiesel2017}, namely,
\Ni
based on independent manual review of edits,
\Nii
based on exploiting community feedback about edits, and
\Niii
based on comparing item states.
As expected, there is a trade-off between corpus size, label quality, and annotation costs. Below, we review state-of-the art approaches for constructing vandalism corpora under each strategy.

\paragraph{Annotation Based on Independent Manual Review}

The most reliable approach to construct a vandalism corpus is to manually review and annotate its edits. When facing millions of edits, however, the costs for a manual review become prohibitive, thus severely limiting corpus size. The largest manually annotated vandalism corpus to date is the PAN Wikipedia Vandalism Corpus~2010 and~2011, comprising a sample of 30,000~edits each of which having manually been annotated via crowdsourcing using Amazon's Mechanical Turk~\citep{Potthast2010_Crowdsourcing}. About 7\%~of the edits have been found to be vandalism. This approach, however, is probably not suited to Wikidata: an average worker on Mechanical Turk is much less familiar with Wikidata, and the expected ratio of vandalism in a random sample of Wikidata edits is about~0.2\% (compared with~7\% in Wikipedia), so that a significantly higher number of edits would have to be reviewed in order to obtain a sensible number of vandalism cases for training a model.

\paragraph{Annotation Based on Community Feedback}

A more scalable approach to construct a vandalism corpus is to rely on feedback about edits provided by the community for annotations. However, not all edits made to Wikidata are currently reviewed by the community, thus limiting the recall in a sample of edits to the amount of vandalism that is actually caught, and, not all edits that are rolled back are true vandalism. Nevertheless, for its simplicity, this approach was adopted to construct the Wikidata Vandalism Corpus (WDVC)~2015~\citep{Heindorf2015} and~2016, whereas the latter was employed as evaluation corpus at the WSDM Cup~2017. Both corpus versions are freely available for download.%
\footnote{\scriptsize See \url{http://www.wsdm-cup-2017.org/vandalism-detection.html}}
The corpus construction is straightforward: based on the portion of the Wikidata dump with manually created revisions, those revisions that have been reverted via Wikidata's rollback facility are labeled vandalism. The rollback facility is a special instrument to revert vandalism; it is accessible to privileged Wikidata editors only. This makes our corpus robust against manipulation by non-privileged and, in particular, anonymous Wikidata editors. As a result, we obtain a large-scale vandalism corpus comprised of more than 82~million manual revisions that were made between October~2012 and June~2016. About 200,000~revisions have been rolled back in this time (and hence are labeled vandalism). By a manual analysis we got evidence that~86\% of the revisions labeled vandalism are indeed vandalism as per Wikidata's definition~\citep{Heindorf2015}.%
\footnote{\scriptsize\url{https://www.wikidata.org/wiki/Wikidata:Vandalism}}
Recently, vandalism corpora for Wikipedia have also been constructed based on community feedback.

\citet{Tran2013,Tran2015_Cross-Language} and \citet{Tran2015_Context-Aware} label all revisions with certain keywords in the revision comment as vandalism, e.g., `vandal' or `rvv' (revert due to vandalism), based on \citeauthor{Kittur2007}'s~\citep{Kittur2007} approach to identify vandalism. These keywords do not work for Wikidata, since revision comments are almost always automatically generated and cannot be changed by editors.

\paragraph{Annotation Based on Item State}

An alternative (and still scalable) approach to build a vandalism corpus is to analyze recurring item states in order to identify so-called item ''reverts'' to previous states in the respective item history. In addition to community feedback this approach does also consider all other events that may have caused an item state to reappear, e.g., in case that someone just fixes an error without noticing that the error was due to vandalism. As a consequence, a higher recall can be achieved whereas, however, a lower precision must be expected. \citet{Sarabadani2017} adopted this approach, and, in order to increase precision, they suggested a set of rules to annotate an edit as vandalism if and only if
\Ni
it is a manual edit,
\Nii
it has been reverted,
\Niii
it does not originate from a pre-defined privileged editor group,%
\footnote{\scriptsize These include the groups sysop, checkuser, flood, ipblock-exempt, oversight, prop\-erty-creator, rollbacker, steward, sysop, translationadmin, and wikidata-staff.}
\Niv
it has not been propagated from a dependent Wikipedia project, and
\Nv
it does not merge two Wikidata items.
For unknown reasons, not all edits of Wikidata's history are annotated this way but only a subset of 500,000 in~2015, yielding altogether only about 20,000 vandalism edits. While the authors claim superiority of the design of their corpus over ours, their self-reported precision values are not convincing: while only~68\% of the edits labeled vandalism are in fact vandalism (86\% in our corpus), 92\%~of the edits are reported to be at least "damaging'' to a greater or lesser extent. The authors have reviewed only 100 edits to substantiate these numbers (we have reviewed 1,000 edits), so that these numbers must be taken with a grain of salt.

Altogether, both corpora are suboptimal with regard to recall: within both corpora, about~1\% of the edits are wrongly labeled non-vandalism, which currently amounts to an estimated 800,000 missed vandalism edits over Wikidata's entire history. A machine learning approach to vandalism detection must hence be especially robust against false negatives in the training dataset.

\citet{Tan2014} compiled a dataset of low-quality triples in Freebase according to the following heuristics: Additions that have been deleted within 4~weeks after their submission are considered low-quality, as well as removals that have not been reinserted within 4~weeks. The manual investigation of 200~triples revealed a precision of about~89\%. However, the usefulness of the Freebase data set is restricted by the fact that Google has shut down Freebase; the Freebase data is currently being transferred to Wikidata~\citep{Pellissier2016}.

\subsection{Vandalism Detection}

The detection of vandalism and damaging edits in \emph{structured} knowledge bases such as triple stores is a new research area. Hence, only three approaches have been published before the WSDM Cup~2017, which represent the state of the art~\citep{Heindorf2016,Sarabadani2017,Tan2014}. All employ machine learning, using features derived from both an edit's content and its context. In what follows, we briefly review them.

The most effective approach to Wikidata vandalism detection, WDVD, was proposed by \citet{Heindorf2016}. It implements 47~features, from which 27 encode an edit's {\em content} and 20 an edit's {\em context}. The content-based features cover character level features, word level features, and sentence-level features, all are computed from the automatically generated revision comment. In addition, WDVD employs features to capture predicates and objects of subject-predicate-object triples. Context-based features include user reputation, user geolocation, item reputation, and the meta data of an edit. As classification algorithm random forests along with multiple-instance learning is employed. Multiple-instance learning is applied to consecutive edits by the same user on the same item, so-called editing sessions. Typically, editing sessions are closely related, so that multiple-instance learning has a significant positive effect on the classification performance.

The second-most effective approach has been deployed within Wikimedia's Objective Revision Evaluation Service, ORES~\citep{Sarabadani2017}. ORES operationalizes 14 features (see Table~\ref{table-features}), most of which were introduced with WDVD, since the WDVD developers shared with Wikimedia a detailed feature list. Meanwhile, certain features have been discarded from WDVD due to overfitting but are still found in the ORES system. Altogether, the effectiveness reported by \citeauthor{Sarabadani2017} is significantly worse compared with WDVD~\citep{Heindorf2016}.

\citet{Tan2014} developed a classifier to detect low-quality contributions to Freebase. The only content-based features are the predicates of subject-predicate-object triples, which are used to predict low-quality contributions. Regarding context-based features, the developers employ user history features including numbers of past edits (total, correct, incorrect) and the age of the user account. Also user expertise is captured by representing users in a topic space based on their past contributions: a new contribution is mapped into the topic space and compared to the user's past revisions using the dot product, the cosine similarity, and the Jaccard index. None of the existing approaches has evaluated so far user expertise features for Wikidata.

The three approaches above build upon previous work on {\em Wikipedia} vandalism detection, whereas the first machine learning-based approach for this task was proposed by \citet{Potthast2008}. It was based on 16~features, primarily focusing on content, detecting nonsense character sequences and vulgarity. In subsequent work, and as a result of two shared task competitions at PAN~2010 and PAN~2011~\citep{Potthast2011_Overview,Potthast2010_Overview}, the original feature set was substantially extended; \citet{Adler2011} integrated many of them. From the large set of approaches that have been proposed since then, those of \citet{Wang2010} and \citet{Ramaswamy2013} stick out: they are based on search engines to check the correctness of Wikipedia edits, achieving a better performance than previous approaches. On the downside, their approaches have only been evaluated on a small dataset and cannot be easily scaled-up to hundreds of millions of edits. To improve the context-based analysis it was proposed to measure user reputation~\citep{Adler2010} as well as spatio-temporal user information~\citep{West2010}. Again, \citet{Kumar2015} stick out since they do not try to detect damaging edits but vandalizing users via their behavior. Many of these features have been transferred to Wikidata vandalism detection; however, more work will be necessary to achieve detection performance comparable to Wikipedia vandalism detectors.

\enlargethispage{\baselineskip}

\section{Evaluation Framework}
\label{evaluation-framework}

This section introduces the knowledge base Wikidata in brief,%
\footnote{\scriptsize See \url{https://www.wikidata.org/wiki/Help:Contents} for a comprehensive overview.}
the Wikidata vandalism corpus derived from it, the evaluation platform, the used performance measures, as well as the baselines.

\subsection{Wikidata}

Wikidata is organized around items. Each item describes a coherent concept from the real world, such as a person, a city, an event, etc. An item in turn can be divided into an item head and an item body. The item head consists of human-readable labels, descriptions, and aliases, provided for up to 375 supported language codes. The item body consists of structured statements, such as the date of birth of a person, as well as sitelinks to Wikipedia pages that cover the same topic as the item. Each time a user edits an item, a new revision is created within the item's revision history. We refer to consecutive revisions from the same user on the same item as an ``editing session''.

\subsection{Wikidata Vandalism Corpus 2016}
\label{subsec:corpus}

For the shared task we built the Wikidata Vandalism Corpus~2016 (short: WDVC-2016%
\footnote{\scriptsize Corpus available at \url{http://www.wsdm-cup-2017.org/vandalism-detection.html}.}),
which is an updated version of the WDVC-2015 corpus~\citep{Heindorf2015}. The corpus consists of 82~million user-contributed revisions made between October~2012 to June~2016 (excluding revisions from bots) alongside 198,147 vandalism annotations on those revisions that have been reverted via the administrative rollback feature; the feature is employed at Wikidata with the purpose to revert vandalism and similarly damaging contributions. Moreover, our corpus provides meta information that is not readily available from Wikidata, such as geolocation data of all anonymous edits as well as Wikidata revision tags originating from both the Wikidata Abuse Filter and semi-automatic editing tools. Table~\ref{table-evaluation-datasets} gives an overview of the corpus: participants of the shared task were provided training data and validation data, while the test data was held back until after the final evaluation.

\begin{table}[tb]
\renewcommand{\arraystretch}{1.0}

\scriptsize
\centering
\setlength{\tabcolsep}{2.75pt}

\caption{Datasets for training, validation, and test in terms of time period covered, vandalism revisions, total revisions, sessions, items, and users. Numbers are given in thousands.}%
\label{table-evaluation-datasets}

\begin{tabular}{@{}lrrrrrrrr@{}}
\addlinespace

\toprule
\bfseries Dataset & \multicolumn{1}{c}{\bfseries From} & \multicolumn{1}{c}{\bfseries To} & \bfseries Vand. & \bfseries Rev. & \bfseries Sessions & \bfseries Items & \bfseries Users \\ 
\midrule
Training   & Oct 1, 2012 & Feb 29, 2016 & 176 & 65,010 & 36,552 & 12,401 & 471 \\ 
Validation & Mar 1, 2016 & Apr 30, 2016 &  11 &  7,225 &  3,827 &  3,116 &  43 \\ 
\midrule
Test       & May 1, 2016 & Jun 30, 2016 &  11 & 10,445 &  3,122 &  2,661 &  41 \\ 
\bottomrule

\end{tabular}

\setlength{\tabcolsep}{2.25pt}

\vspace{3ex}
\caption{The Wikidata Vandalism Corpus WDVC-2016 in terms of total unique users, items, sessions, and revisions with a breakdown by item part and by vandal(ism) status (Vand.). Numbers are given in thousands.}
\label{table-corpus-statistics}

\begin{tabular}{@{}lrrrrrrrrr@{}}
\addlinespace

\toprule
& \multicolumn{3}{c}{\bfseries Entire corpus} & \multicolumn{3}{c}{\bfseries Item head} & \multicolumn{3}{c}{\bfseries Item body} \\
\cmidrule(l@{\tabcolsep}r@{\tabcolsep}){2-4}\cmidrule(l@{\tabcolsep}r@{\tabcolsep}){5-7}\cmidrule(l@{\tabcolsep}){8-10}
&  Total & Vand. & Regular & Total & Vand. & Regular &  Total & Vand. & Regular \\
\midrule
Revisions &   82,680 &       198 &  82,482 &    16,602 &       100 &  16,502 &    59,699 &        92 &  59,606 \\
Sessions  &   43,254 &       119 &  43,142 &     9,835 &        71 &   9,765 &    33,955 &        49 &  33,908 \\
Items     &   14,049 &        85 &  14,049 &     6,268 &        54 &   6,254 &    12,744 &        39 &  12,741 \\ 
Users     &      518 &        96 &     431 &       247 &        65 &     186 &       310 &        35 &     279 \\  
\bottomrule 

\end{tabular}
\end{table}

\begin{figure}[tb]
\includegraphics[width=\columnwidth]{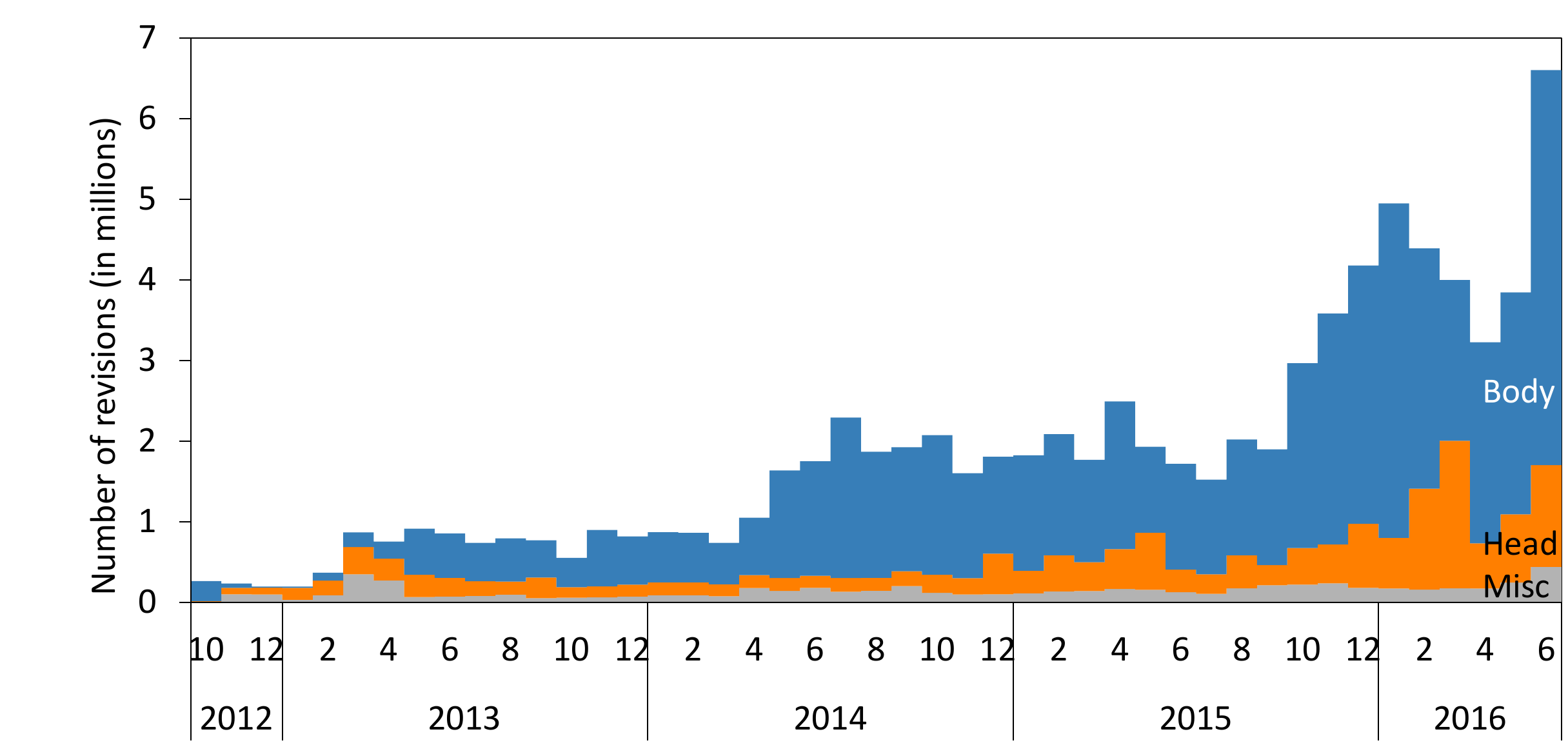}
\includegraphics[width=\columnwidth]{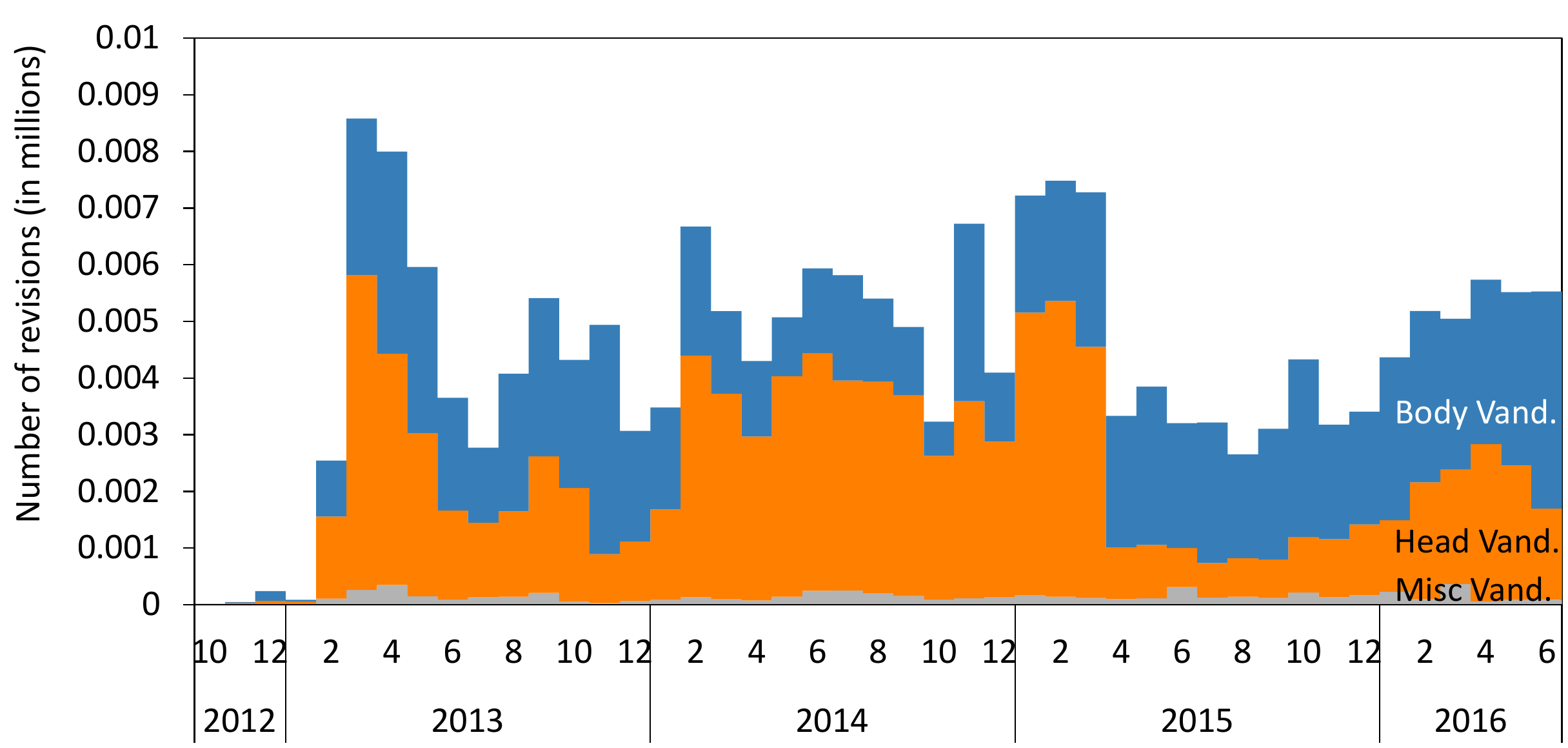}
\caption{Overview of the Wikidata Vandalism Corpus 2016 by content type: Number of {\em all} revisions by content type (top) and number of {\em vandalism} revisions by content type (bottom).}\label{plot-corpus-over-time}
\vspace{-1ex}
\end{figure}

Table~\ref{table-corpus-statistics} gives an overview about the corpus in terms of content type (head vs.\ body) as well as revisions, sessions, items, and users. Figure~\ref{plot-corpus-over-time} plots the development of the corpus over time. While the number of revisions per month is increasing (top), the number of vandalism revisions per month varies without a clear trend (bottom). We attribute the observed variations to the fast pace at which Wikidata is developed, both in terms of data acquisition and frontend development. For example, the drop in vandalism affecting item heads around April~2015 is probably related to the redesign of Wikidata's user interface around this time: with the new user interface it is less obvious to edit labels, descriptions, and aliases which might deter many drive-by vandals.%
\footnote{\scriptsize \url{https://lists.wikimedia.org/pipermail/wikidata/2015-March/005703.html}}

\enlargethispage{\baselineskip}
\subsection{Evaluation Platform}

Our evaluation platform has been built to ensure
\Ni
reproducibility of results,
\Nii
blind evaluation,
\Niii
ground truth protection, and,
\Niv
to implement a realistic scenario of vandalism detection where revisions are scored in near real-time as they arrive.
Reproducibility is ensured by inviting participants to submit their software instead of just its run output, employing the evaluation-as-a-service platform TIRA~\citep{Potthast2014}. TIRA implements a cloud-based evaluation system, where each participant gets assigned a virtual machine in which a working version of their approach is to be deployed. The virtual machines along with the deployed softwares are remote-controlled via TIRA's web interface. Once participants manage to run their software on the test dataset hosted on TIRA, the virtual machine can be archived for later re-evaluation. In addition, TIRA ensures that participants do not get direct access to the test datasets, giving rise to blind evaluation, where the to-be-evaluated software's authors have never observed the test datasets directly.

For following reasons the task of vandalism detection in Wikis is intrinsically difficult to be organized as a shared task: the ground truth is publicly available via Wikimedia's data dumps, and the vandalism occurring at some point in the revision history of a Wikidata item will eventually be undone manually via the rollback facility, which we also use to create our ground truth. Handing out a test dataset spanning a long time frame would therefore effectively reveal the ground truth to participants. At the same time, in practice, it would be rather an unusual task to be handed a large set of consecutive revisions to be classified as vandalism or no, when in fact, every revision should be reviewed as it arrives. We therefore opted for a streaming-based setup where the participant software was supposed to connect to a data server,%
\footnote{\scriptsize \url{https://github.com/wsdm-cup-2017/wsdmcup17-data-server}}
which initiates a stream of revisions in chronological order (by increasing revision IDs), but waits sending new ones unless the software has returned vandalism scores for those already sent. This way, a software cannot easily exploit information gleaned from revisions occurring after a to-be-classified revision. However, we did not expect synchronous classification, but allowed for a backpressure window of $k=16$ revisions, so that revision $i + k$ is sent as soon as the score for $i$-th revision has been reported. This allows for concurrent processing of data and the analysis, while preventing a deep look-ahead into the future of a revision. At $k=16$, no more than 0.003\% of the ground truth was revealed within the backpressure window (283 regular revisions and 76 vandalism revisions). However, as reported by participants, vandalism scores were based on previous revisions, while the backpressure window was unanimously used to gain computational speedup, and not to exploit ground truth or editing sessions.

\subsection{Performance Measures}

We employ the same performance measures as in our previous work \citep{Heindorf2016}: area under curve of the receiver operating characteristic, \rocauc, and area under the precision-recall curve, \prauc. Both \rocauc and \prauc cover a wide range of operating points, emphasizing different characteristics. Given the large class imbalance of vandalism to non-vandalism, \rocauc emphasizes performance in the high recall range, while \prauc emphasizes performance in the high precision range. I.e., \rocauc is probably more meaningful for a semi-automatic operation of a vandalism detector, pre-filtering revisions likely to be vandalism and leaving the final judgment to human reviewers. \prauc addresses scenarios where vandalism shall be reverted fully automatically without any human intervention. The winner of the WSDM Cup was determined based on \rocauc.

For informational purposes, we also report the typical classifier performance measures for the operating point at the threshold~0.5: accuracy (ACC), precision (P), recall (R), $F_1$ measure (F).

\subsection{Baselines and Meta Approach}

\bslabel{WDVD}
We employ the Wikidata Vandalism Detector, WDVD\citep{Heindorf2016}, as (strong) state-of-the-art baseline. The underlying model consists of 47~features and employs multiple-instance learning on top of bagging and random forests. The model was trained on training data ranging from May~1, 2013, to April~30, 2016. We used the same hyperparameters for this model as reported in our previous work~\citep{Heindorf2016}: 16~random forests, each build on 1/16~of the training dataset with the forests consisting of 8~trees, each having a maximal depth of~32 with two features per split and using the default Gini split criterion. In order to adjust WDVD to the new evaluation setup where revisions arrive constantly in a stream, we adjusted the multiple-instance learning to consider only those revisions of a session up until the current revision.


\bslabel{FILTER}
As a second baseline, we employ so-called revision tags, which are created on Wikidata due to two main mechanisms:
\Ni
The Wikidata Abuse Filter automatically tags revisions according to a collection of human-generated rules.
\Nii
Revisions created by semi-automatic editing tools such as the Wikidata Game are tagged with the authentication method used by the semi-automatic editing tool.
In general, tags assigned by the abuse filter are a strong signal for vandalism while tags from semi-automatic editing tools are a signal for non-vandalism. We trained a random forest with scikit-learn's default hyperparameters on the training data from May~1, 2013 to April~30, 2016. The revision tags were provided to all participants as part of the meta data. This baseline has also been incorporated into WDVD as feature \texttt{revisionTags}.

\bslabel{ORES}
We reimplemented the ORES approach~\citep{Sarabadani2017} developed for Wikidata vandalism detection by the Wikimedia Foundation and we apply it to the Wikidata Vandalism Corpus~2016.%
\footnote{\scriptsize Compared with our previous paper~\citep{Heindorf2016}, we employ an updated version of ORES that was recently published by~\citet{Sarabadani2017}.}
Essentially, this approach consists of a subset of WDVD's features plus some additional features that were previously found to overfit~\citep{Heindorf2016}. It uses a random forest and was trained on the training data ranging from May~1, 2013 to April~30, 2016. We use the original hyperparameters by \citeauthor{Sarabadani2017}:%
\footnote{\scriptsize \url{https://github.com/wiki-ai/wb-vandalism/blob/sample_subsets/Makefile}}
80 decision trees considering `log2' features per split using the `entropy' criterion.\footnote{\scriptsize The difference in performance of our reimplementation of ORES compared with \citet{Sarabadani2017} is explained by the different datasets and evaluation metrics: while we split the dataset by time, \citeauthor{Sarabadani2017} split the dataset randomly, causing revisions from the same editing session to appear both in the training as well as test dataset.} While \citet{Sarabadani2017} experimented with balancing the weights of the training examples, we do not do so for the ORES baseline since it has no effect on performance in terms of \rocauc and decreases performance in terms of \prauc.

\bslabel{META}
Given the vandalism scores returned by participant approaches and our baselines, the question arises what the detection performance would be if all these approaches were combined into one. To get an estimation of the possible performance, we employ a simple meta approach whose score for each to-be-classified revision corresponds to the mean of all 8~approaches. As it turns out, the meta approach slightly outperforms the other approaches.

\providecommand{\bscolorcell}[6]{}
\renewcommand{\bscolorcell}[6]{%
  \fcolorbox{#1}{#3}{\parbox{#4}{\centering\vspace{-2pt}\color{#2}\rule[-0pt]{-2pt}{#5}\kern-1pt#6\kern-2.25pt\vspace{-4pt}}}}

\providecommand{\bscellA}[2][]{}
\renewcommand{\bscellA}[2][]{%
  \providecommand{\grayfg}{0.00}%
  \providecommand{\graybg}{1.00}%
  \ifnum #2 = 0  \renewcommand{\grayfg}{0.00}\renewcommand{\graybg}{1.00}\else\fi%
  \ifnum #2 > 0  \renewcommand{\grayfg}{0.00}\renewcommand{\graybg}{0.86}\else\fi%
  \ifnum #2 > 5  \renewcommand{\grayfg}{0.00}\renewcommand{\graybg}{0.82}\else\fi%
  \ifnum #2 > 10 \renewcommand{\grayfg}{0.00}\renewcommand{\graybg}{0.78}\else\fi%
  \ifnum #2 > 15 \renewcommand{\grayfg}{0.00}\renewcommand{\graybg}{0.74}\else\fi%
  \ifnum #2 > 20 \renewcommand{\grayfg}{0.00}\renewcommand{\graybg}{0.70}\else\fi%
  \ifnum #2 > 25 \renewcommand{\grayfg}{0.00}\renewcommand{\graybg}{0.66}\else\fi%
  \ifnum #2 > 30 \renewcommand{\grayfg}{0.00}\renewcommand{\graybg}{0.62}\else\fi%
  \ifnum #2 > 35 \renewcommand{\grayfg}{0.00}\renewcommand{\graybg}{0.58}\else\fi%
  \ifnum #2 > 40 \renewcommand{\grayfg}{0.00}\renewcommand{\graybg}{0.54}\else\fi%
  \ifnum #2 > 45 \renewcommand{\grayfg}{0.00}\renewcommand{\graybg}{0.50}\else\fi%
  \ifnum #2 > 50 \renewcommand{\grayfg}{0.00}\renewcommand{\graybg}{0.46}\else\fi%
  \ifnum #2 > 55 \renewcommand{\grayfg}{1.00}\renewcommand{\graybg}{0.42}\else\fi%
  \ifnum #2 > 60 \renewcommand{\grayfg}{1.00}\renewcommand{\graybg}{0.38}\else\fi%
  \ifnum #2 > 65 \renewcommand{\grayfg}{1.00}\renewcommand{\graybg}{0.34}\else\fi%
  \ifnum #2 > 70 \renewcommand{\grayfg}{1.00}\renewcommand{\graybg}{0.30}\else\fi%
  \ifnum #2 > 75 \renewcommand{\grayfg}{1.00}\renewcommand{\graybg}{0.26}\else\fi%
  \ifnum #2 > 80 \renewcommand{\grayfg}{1.00}\renewcommand{\graybg}{0.22}\else\fi%
  \ifnum #2 > 85 \renewcommand{\grayfg}{1.00}\renewcommand{\graybg}{0.18}\else\fi%
  \ifnum #2 > 90 \renewcommand{\grayfg}{1.00}\renewcommand{\graybg}{0.14}\else\fi%
  \ifnum #2 > 95 \renewcommand{\grayfg}{1.00}\renewcommand{\graybg}{0.10}\else\fi%
  \definecolor{cellfg}{gray}{\grayfg}%
  \definecolor{cellbg}{gray}{\graybg}%
  \bscolorcell{white}{cellfg}{cellbg}{2em}{1ex}{#1}}

\providecommand{\bscellB}[2][]{}
\renewcommand{\bscellB}[2][]{%
  \providecommand{\grayfg}{0.00}%
  \providecommand{\graybg}{1.00}%
  \ifnum #2 = 0  \renewcommand{\grayfg}{0.00}\renewcommand{\graybg}{1.00}\else\fi%
  \ifnum #2 > 0  \renewcommand{\grayfg}{0.00}\renewcommand{\graybg}{1.00}\else\fi%
  \ifnum #2 > 5  \renewcommand{\grayfg}{0.00}\renewcommand{\graybg}{1.00}\else\fi%
  \ifnum #2 > 10 \renewcommand{\grayfg}{0.00}\renewcommand{\graybg}{1.00}\else\fi%
  \ifnum #2 > 15 \renewcommand{\grayfg}{0.00}\renewcommand{\graybg}{1.00}\else\fi%
  \ifnum #2 > 20 \renewcommand{\grayfg}{0.00}\renewcommand{\graybg}{1.00}\else\fi%
  \ifnum #2 > 25 \renewcommand{\grayfg}{0.00}\renewcommand{\graybg}{1.00}\else\fi%
  \ifnum #2 > 30 \renewcommand{\grayfg}{0.00}\renewcommand{\graybg}{1.00}\else\fi%
  \ifnum #2 > 35 \renewcommand{\grayfg}{0.00}\renewcommand{\graybg}{1.00}\else\fi%
  \ifnum #2 > 40 \renewcommand{\grayfg}{0.00}\renewcommand{\graybg}{1.00}\else\fi%
  \ifnum #2 > 45 \renewcommand{\grayfg}{0.00}\renewcommand{\graybg}{1.00}\else\fi%
  \ifnum #2 > 50 \renewcommand{\grayfg}{0.00}\renewcommand{\graybg}{1.00}\else\fi%
  \ifnum #2 > 55 \renewcommand{\grayfg}{0.00}\renewcommand{\graybg}{0.82}\else\fi%
  \ifnum #2 > 60 \renewcommand{\grayfg}{0.00}\renewcommand{\graybg}{0.74}\else\fi%
  \ifnum #2 > 65 \renewcommand{\grayfg}{0.00}\renewcommand{\graybg}{0.66}\else\fi%
  \ifnum #2 > 70 \renewcommand{\grayfg}{0.00}\renewcommand{\graybg}{0.58}\else\fi%
  \ifnum #2 > 75 \renewcommand{\grayfg}{1.00}\renewcommand{\graybg}{0.50}\else\fi%
  \ifnum #2 > 80 \renewcommand{\grayfg}{1.00}\renewcommand{\graybg}{0.42}\else\fi%
  \ifnum #2 > 85 \renewcommand{\grayfg}{1.00}\renewcommand{\graybg}{0.34}\else\fi%
  \ifnum #2 > 90 \renewcommand{\grayfg}{1.00}\renewcommand{\graybg}{0.26}\else\fi%
  \ifnum #2 > 95 \renewcommand{\grayfg}{1.00}\renewcommand{\graybg}{0.14}\else\fi%
  \definecolor{cellfg}{gray}{\grayfg}%
  \definecolor{cellbg}{gray}{\graybg}%
  \bscolorcell{white}{cellfg}{cellbg}{2em}{1ex}{#1}}

\providecommand\bsrot{}
\providecommand\OK{}
\providecommand\no{}
\providecommand\sm{}

\renewcommand{\bsrot}[1]{\kern-0.5em\rotatebox{40}{#1}\kern-0.5em}
\renewcommand*\OK{\ding{51}}
\renewcommand*\no{--}
\renewcommand*\sm{(\ding{51})}

\begin{table*}

\centering
\fontsize{7.0}{7.0}\selectfont
\setlength{\tabcolsep}{6.45pt}
\renewcommand{\arraystretch}{0.96}

\caption{Overview of feature groups, features, and their usage by WSDM Cup participants. Features that were newly introduced by participants and not previously used as part of WDVD are marked with an astersik (*). Features computed in the same way as for WDVD as well as new features are marked with \OK, features computed similarly to WDVD are marked with \sm, features for which it is unclear from their respective paper whether they are included are marked with ?, and features not utilized are marked with \no.}
\label{table-features}

\begin{tabular}{@{}lrl@{\hspace{3em}}cccccccc@{}}
\addlinespace

\toprule
\multicolumn{2}{@{}l}{\bfseries Feature group} & \bfseries Feature & \multicolumn{5}{@{}c@{}}{\bfseries Submitted Approach} & \multicolumn{3}{@{\hspace{\tabcolsep}}c@{}}{\bfseries Baseline} \\
\cmidrule(r@{\tabcolsep}){1-2}\cmidrule(l@{\tabcolsep}r@{3em}){3-3}\cmidrule{4-8}\cmidrule(l@{\tabcolsep}){9-11}
&&& \bsrot{Buffaloberry~\citep{Crescenzi2017}} & \bsrot{Conkerberry~\citep{Grigorev2017}} & \bsrot{Honeyberry~\citep{Yamazaki2017}} & \bsrot{Loganberry~\citep{Zhu2017}} & \bsrot{Riberry~\citep{Yu2017}} & \bsrot{WDVD~\citep{Heindorf2016}} & \bsrot{FILTER~\citep{Heindorf2016}} & \bsrot{ORES~\citep{Sarabadani2017}} \\

\midrule
\multirow{35}{*}{\rotatebox[origin=c]{90}{\normalsize Content features}}

&  \kern-0.5em Character features
&  \lowerCaseRatio                 & \OK & \no & \OK & \OK & \OK & \OK & \no & \no \\
&& \upperCaseRatio                 & \OK & \no & \OK & \OK & \OK & \OK & \no & \no \\
&& \nonLatinRatio                  & \OK & \no & \OK & \OK & \OK & \OK & \no & \no \\
&& \latinRatio                     & \OK & \no & \OK & \OK & \OK & \OK & \no & \no \\
&& \alphanumericRatio              & \OK & \no & \OK & \OK & \OK & \OK & \no & \no \\
&& \digitRatio                     & \OK & \no & \OK & \OK & \OK & \OK & \no & \no \\
&& \punctuationRatio               & \OK & \no & \OK & \OK & \OK & \OK & \no & \no \\
&& \whitespaceRatio                & \OK & \no & \OK & \OK & \OK & \OK & \no & \no \\
&& \longestCharacterSequence       & \OK & \no & \OK & \OK & \OK & \OK & \no & \no \\
&& \asciiRatio                     & \no & \no & \no & \OK & \OK & \OK & \no & \no \\
&& \bracketRatio                   & \OK & \no & \no & \OK & \OK & \OK & \no & \no \\
&& misc features from WDVD         & \no & \no & \no & \no & \OK & \no & \no & \no \\
&& \symbolRatio*                   & \OK & \no & \no & \no & \no & \no & \no & \no \\
&& \mainAlphabet*                  & \OK & \no & \no & \no & \no & \no & \no & \no \\
                                                     
&  \kern-0.5em Word features                         
&  \languageWordRatio              & \OK & \no & \OK & \OK & \OK & \OK & \no & \no \\
&& \containsLanguageWord           & \OK & \no & \OK & \OK & \OK & \OK & \no & \no \\
&& \lowerCaseWordRatio             & \OK & \no & \OK & \OK & \OK & \OK & \no & \no \\
&& \longestWord                    & \OK & \no & \OK & \OK & \OK & \OK & \no & \no \\
&& \containsURL                    & \OK & \no & \OK & \OK & \OK & \OK & \no & \no \\
&& \badWordRatio                   & \OK & \no & \no & \OK & \OK & \OK & \no & \no \\
&& \proportionOfQidAdded           & \no & \no & \no &   ? & \no & \OK & \no & \OK \\
&& \upperCaseWordRatio             & \OK & \no & \OK & \OK & \OK & \OK & \no & \no \\
&& \proportionOfLinksAdded         & \no & \no & \no &  ?  & \no & \OK & \no & \OK \\
&& \proportionOfLanguageAdded      & \no & \no & \no & \no & \no & \no & \no & \OK \\
&& misc features from WDVD         & \no & \no & \no & \no & \OK & \no & \no & \no \\
&& \bagOfWords*                    & \no & \OK & \no & \no & \no & \no & \no & \no \\
                                                     
&  \kern-0.5em Sentence features                     
&  \commentTailLength              & \OK & \no & \no & \OK & \OK & \OK & \no & \no \\
&& \commentSitelinkSimilarity      & \sm & \no & \no & \OK & \OK & \OK & \no & \no \\
&& \commentLabelSimilarity         & \sm & \no & \no & \OK & \OK & \OK & \no & \no \\
&& \commentCommentSimilarity       & \no & \no & \no &  ?  & \no & \OK & \no & \no \\
&& \languageMatchProb*             & \OK & \no & \no & \no & \no & \no & \no & \no \\
&& \hasIdentifierChanged           & \no & \no & \no & \no & \no & \no & \no & \OK \\
                                                     
&  \kern-0.5em Statement features                    
&  \propertyFrequency              & \no & \sm & \no & \OK & \OK & \OK & \no & \sm \\
&& \itemValueFrequency             & \no & \sm & \no & \OK & \OK & \OK & \no & \no \\
&& \literalValueFrequency          & \no & \sm & \no & \OK & \OK & \OK & \no & \no \\

\midrule
\multirow{45}{*}{\rotatebox[origin=c]{90}{\normalsize Context features}}

&  \kern-0.5em User features
&  \userCountry                    & \OK & \OK & \OK & \no & \OK & \OK & \no & \no \\
&& \userTimeZone                   & \OK & \OK & \OK & \no & \OK & \OK & \no & \no \\
&& \userCity                       & \OK & \OK & \OK & \no & \OK & \OK & \no & \no \\
&& \userCounty                     & \OK & \OK & \OK & \no & \OK & \OK & \no & \no \\
&& \userRegion                     & \OK & \OK & \OK & \no & \OK & \OK & \no & \no \\
&& \cumUserUniqueItems             & \no & \no & \no & \no & \OK & \OK & \no & \no \\
&& \userContinent                  & \OK & \OK & \OK & \no & \OK & \OK & \no & \no \\
&& \isRegisteredUser               & \OK & \OK & \OK & \OK & \OK & \OK & \no & \OK \\
&& \userFrequency                  & \OK & \OK & \no & \no & \OK & \OK & \no & \no \\
&& \isPrivilegedUser               & \OK & \no & \no & \OK & \OK & \OK & \no & \sm \\
&& misc features from WDVD         & \no & \no & \no & \no & \OK & \no & \no & \no \\
&& \userIPSubnets*                 & \no & \OK & \no & \no & \no & \no & \no & \no \\
&& \userVandalismFraction*         & \no & \no & \no & \OK & \OK & \no & \no & \no \\
&& \userVandalismCount*            & \no & \no & \no & \OK & \no & \no & \no & \no \\
&& \userUniqueItems*               & \no & \no & \no & \no & \OK & \no & \no & \no \\
&& \userAge                        & \no & \no & \no & \no & \no & \no & \no & \OK \\
                                                     
&  \kern-0.5em Item features                         
&  \logCumItemUniqueUsers          & \no & \no & \no & \no & \no & \OK & \no & \no \\
&& \logItemFrequency               & \no & \no & \no & \no & \no & \OK & \no & \no \\
&& \isHuman                        & \no & \no & \no & \no & \no & \no & \no & \OK \\
&& \isLivingPerson                 & \no & \no & \no & \no & \no & \no & \no & \OK \\
&& misc features from WDVD         & \no & \no & \no & \no & \OK & \no & \no & \no \\
&& \itemFrequency*                 & \OK & \sm & \no & \no & \OK & \no & \no & \no \\
&& \itemVandalismFraction*         & \no & \no & \no & \OK & \OK & \no & \no & \no \\
&& \itemVandalismCount*            & \no & \no & \no & \OK & \no & \no & \no & \no \\
&& \itemUniqueUsers*               & \no & \no & \no & \no & \OK & \no & \no & \no \\
                                                     
&  \kern-0.5em Revision features                     
&  \revisionTags                   & \sm & \OK & \OK &   ? & \OK & \OK & \OK & \no \\
&& \revisionLanguage               & \sm & \OK & \OK & \OK & \OK & \OK & \no & \no \\
&& \revisionAction                 & \OK & \OK & \OK & \OK & \OK & \OK & \no & \sm \\
&& \commentLength                  & \no & \no & \OK & \OK & \OK & \OK & \no & \no \\
&& \isLatinLanguage                & \no & \no & \OK & \OK & \OK & \OK & \no & \no \\
&& \revisionPrevAction             & \no & \no & \no &   ? & \no & \OK & \no & \no \\
&& \revisionSubaction              & \OK & \OK & \OK & \OK & \OK & \OK & \no & \sm \\
&& \positionWithinSession          & \no & \no & \no &   ? & \no & \OK & \no & \no \\
&& \numberOfIdentifiersChanged     & \no & \no & \no & \no & \no & \no & \no & \OK \\
&& misc features from WDVD         & \no & \no & \no & \no & \OK & \no & \no & \no \\
&& \isMinorRevision*               & \OK & \no & \no & \no & \OK & \no & \no & \no \\
&& \changeCount*                   & \OK & \OK & \no & \no & \OK & \no & \no & \sm \\
&& \superItem*                     & \OK & \no & \no & \no & \OK & \no & \no & \no \\
&& \revisionSize*                  & \OK & \no & \no & \no & \OK & \no & \no & \no \\
&& \hourOfDay*                     & \OK & \no & \no & \no & \no & \no & \no & \no \\
&& \dayOfWeek*                     & \no & \no & \no & \no & \OK & \no & \no & \no \\
&& \revisionPrevUser*              & \OK & \no & \no & \no & \no & \no & \no & \no \\
&& \hashTag*                       & \no & \OK & \sm & \no & \no & \no & \no & \no \\
&& \isSpecialRevision*             & \no & \no & \OK & \no & \no & \no & \no & \no \\

\bottomrule

\end{tabular}
\end{table*}

\section{Survey of Submissions}
\label{survey}

This section surveys the features and learning algorithms employed by the participants. All of them chose to build their own model---which eventually are based on our WDVD approach~\citep{Heindorf2016}, whose code base had been published to ensure reproducibility. On the one hand, the availability of this code base leveled the playing field among participants since it enabled everyone to achieve at least state-of-the-art performance. On the other hand, the availability may have stifled creativity and innovation among participants since all approaches follow a similar direction, and no one investigated different classification paradigms. However, all participants attempted to improve over our approach (which was one of the baselines) by developing new features and experimenting with learning variants.

\subsection{Features}

Table~\ref{table-features} gives a comprehensive overview of the features used in the submitted approaches. The feature set is complete; in particular, it unifies those features that are the same or closely related across participants. The table divides the features into two main groups, content features and context features. The content features in turn are subdivided regarding the granularity level at which characteristics are quantified, whereas the context features are subdivided regarding contextual entities in connection with a to-be-classified revision. Since the features have been extensively described in our previous work and the participant notebooks we omit a detailed description here. Instead, the feature names have been chosen to convey their intended semantics and are in accordance with the corresponding implementations found in our code base. For in-depth information we refer to our paper covering WDVD~\citep{Heindorf2016}, the FILTER baseline, our reimplementation of ORES, as well as the notebook papers submitted by the participants~\citep{Crescenzi2017,Grigorev2017,Yamazaki2017,Yu2017,Zhu2017}.

We would like to point out certain observations that can be gained from the overview: Buffaloberry~\citep{Crescenzi2017} used many of the WDVD features but also contributed a number of additional features on top of that. Conkerberry~\citep{Grigorev2017} used an interesting bag of words model that basically consists of the feature values computed by many of the WDVD features, all taken as words. Loganberry~\citep{Zhu2017} did not exploit the information we provided as meta data, such as geolocation, etc. With two exceptions, Honeyberry~\citep{Yamazaki2017} used almost exclusively WDVD features. Riberry~\citep{Yu2017} used on top of the WDVD features those that we previously found to overfit (denoted as ``misc. features from WDVD'' in the table), which may explain their poor overall performance, corroborating our previous results.

\subsection{Learning Algorithms}

Table~\ref{table-algorithms} overviews the employed learning algorithms and organizes them wrt.\ achieved performance. The best-performing approach by Buffaloberry employs XGBoost and multiple-instance learning. The second-best approach by Conkerberry employs a linear SVM, encoding all WDVD features as a bag of words model. This results in an effectiveness comparable to the WDVD baseline in terms of \rocauc, but not in terms of \prauc. The third approach by Loganberry also employs XGBoost, however, in contrast to the first approach, no multiple-instance learning was conducted. The fourth approach, Honeyberry, created an ensemble of various algorithms following a stacking strategy. In contrast to the first approach, the authors put less emphasis on feature engineering. Their final submission contained a bug reducing the performance of their approach, which was fixed only after the submission deadline. The bugfix caused their performance to jump to an \rocauc of~0.928, thus virtually achieving the third place in the competition. The fifth approach of Riberry performed poorly, probably due to overfitting features. The baselines employ a parameter-optimized random forest.

\providecommand\bsrot{}
\providecommand\bscolsep{}
\providecommand\OK{}
\providecommand\no{}
\providecommand\sm{}

\renewcommand{\bsrot}[1]{\rotatebox{40}{#1}}
\renewcommand{\bscolsep}[1][10]{\hspace{-#1pt}}
\renewcommand\OK{\ding{51}}
\renewcommand\no{--}
\renewcommand\sm{(\ding{51})}

\begin{table}[!t]
\fontsize{8}{9}\selectfont
\centering
\renewcommand{\arraystretch}{1.0}
\setlength{\tabcolsep}{0pt}

\caption{Overview of the employed learning algorithms per submission. The rows are sorted wrt.\ the achieved evaluation scores, starting with the best.}
\label{table-algorithms}

\begin{tabular}{l@{\bscolsep[-17]}l@{\bscolsep[12]}l@{\bscolsep[17]}l@{\bscolsep[37]}l@{\bscolsep[24]}l@{\bscolsep[14]}l@{\bscolsep[0]}l@{\bscolsep[29]}l}
\addlinespace
\toprule
\multicolumn{1}{c}{\bfseries Submission} & \multicolumn{8}{c}{\bfseries Learning Algorithms} \\
\cmidrule(r@{0em}){1-1}\cmidrule(l@{1em}){2-9}
& \bsrot{XGBoost} & \bsrot{Linear SVM} & \bsrot{Logistic Regression} & \bsrot{Random Forest} & \bsrot{Extra Trees} & \bsrot{GBT} & \bsrot{Neural Networks} & \bsrot{Multiple-Instance} \\
\midrule
META              & \OK & \OK & \OK & \OK & \OK & \OK & \OK & \OK \\
Buffaloberry      & \OK & \no & \no & \no & \no & \no & \no & \OK \\
Conkerberry       & \no & \OK & \no & \no & \no & \no & \no & \no \\
WDVD (baseline)   & \no & \no & \no & \OK & \no & \no & \no & \OK \\
Honeyberry        & \no & \no & \OK & \OK & \OK & \OK & \OK & \no \\
Loganberry        & \OK & \no & \no & \no & \no & \no & \no & \no \\
Riberry           & \no & \no & \no & \OK & \no & \OK & \no & \no \\
ORES (baseline)   & \no & \no & \no & \OK & \no & \no & \no & \no \\
FILTER (baseline) & \no & \no & \no & \OK & \no & \no & \no & \no \\
\bottomrule
\end{tabular}

\end{table}

\providecommand{\bscolorcell}[6]{}
\renewcommand{\bscolorcell}[6]{%
  \fcolorbox{#1}{#3}{\parbox{#4}{\centering\vspace{-2pt}\color{#2}\rule[-0pt]{-2pt}{#5}\kern-1pt#6\kern-2.25pt\vspace{-4pt}}}}

\providecommand{\bscellA}[2][]{}
\renewcommand{\bscellA}[2][]{%
  \providecommand{\grayfg}{0.00}%
  \providecommand{\graybg}{1.00}%
  \ifnum #2 = 0  \renewcommand{\grayfg}{0.00}\renewcommand{\graybg}{1.00}\else\fi%
  \ifnum #2 > 0  \renewcommand{\grayfg}{0.00}\renewcommand{\graybg}{0.86}\else\fi%
  \ifnum #2 > 5  \renewcommand{\grayfg}{0.00}\renewcommand{\graybg}{0.82}\else\fi%
  \ifnum #2 > 10 \renewcommand{\grayfg}{0.00}\renewcommand{\graybg}{0.78}\else\fi%
  \ifnum #2 > 15 \renewcommand{\grayfg}{0.00}\renewcommand{\graybg}{0.74}\else\fi%
  \ifnum #2 > 20 \renewcommand{\grayfg}{0.00}\renewcommand{\graybg}{0.70}\else\fi%
  \ifnum #2 > 25 \renewcommand{\grayfg}{0.00}\renewcommand{\graybg}{0.66}\else\fi%
  \ifnum #2 > 30 \renewcommand{\grayfg}{0.00}\renewcommand{\graybg}{0.62}\else\fi%
  \ifnum #2 > 35 \renewcommand{\grayfg}{0.00}\renewcommand{\graybg}{0.58}\else\fi%
  \ifnum #2 > 40 \renewcommand{\grayfg}{0.00}\renewcommand{\graybg}{0.54}\else\fi%
  \ifnum #2 > 45 \renewcommand{\grayfg}{0.00}\renewcommand{\graybg}{0.50}\else\fi%
  \ifnum #2 > 50 \renewcommand{\grayfg}{1.00}\renewcommand{\graybg}{0.46}\else\fi%
  \ifnum #2 > 55 \renewcommand{\grayfg}{1.00}\renewcommand{\graybg}{0.42}\else\fi%
  \ifnum #2 > 60 \renewcommand{\grayfg}{1.00}\renewcommand{\graybg}{0.38}\else\fi%
  \ifnum #2 > 65 \renewcommand{\grayfg}{1.00}\renewcommand{\graybg}{0.34}\else\fi%
  \ifnum #2 > 70 \renewcommand{\grayfg}{1.00}\renewcommand{\graybg}{0.30}\else\fi%
  \ifnum #2 > 75 \renewcommand{\grayfg}{1.00}\renewcommand{\graybg}{0.26}\else\fi%
  \ifnum #2 > 80 \renewcommand{\grayfg}{1.00}\renewcommand{\graybg}{0.22}\else\fi%
  \ifnum #2 > 85 \renewcommand{\grayfg}{1.00}\renewcommand{\graybg}{0.18}\else\fi%
  \ifnum #2 > 90 \renewcommand{\grayfg}{1.00}\renewcommand{\graybg}{0.14}\else\fi%
  \ifnum #2 > 95 \renewcommand{\grayfg}{1.00}\renewcommand{\graybg}{0.10}\else\fi%
  \definecolor{cellfg}{gray}{\grayfg}%
  \definecolor{cellbg}{gray}{\graybg}%
  \bscolorcell{white}{cellfg}{cellbg}{2.25em}{1ex}{#1}}

\providecommand{\bscellB}[2][]{}
\renewcommand{\bscellB}[2][]{%
  \providecommand{\grayfg}{0.00}%
  \providecommand{\graybg}{1.00}%
  \ifnum #2 = 0  \renewcommand{\grayfg}{0.00}\renewcommand{\graybg}{1.00}\else\fi%
  \ifnum #2 > 0  \renewcommand{\grayfg}{0.00}\renewcommand{\graybg}{1.00}\else\fi%
  \ifnum #2 > 5  \renewcommand{\grayfg}{0.00}\renewcommand{\graybg}{1.00}\else\fi%
  \ifnum #2 > 10 \renewcommand{\grayfg}{0.00}\renewcommand{\graybg}{1.00}\else\fi%
  \ifnum #2 > 15 \renewcommand{\grayfg}{0.00}\renewcommand{\graybg}{1.00}\else\fi%
  \ifnum #2 > 20 \renewcommand{\grayfg}{0.00}\renewcommand{\graybg}{1.00}\else\fi%
  \ifnum #2 > 25 \renewcommand{\grayfg}{0.00}\renewcommand{\graybg}{1.00}\else\fi%
  \ifnum #2 > 30 \renewcommand{\grayfg}{0.00}\renewcommand{\graybg}{1.00}\else\fi%
  \ifnum #2 > 35 \renewcommand{\grayfg}{0.00}\renewcommand{\graybg}{1.00}\else\fi%
  \ifnum #2 > 40 \renewcommand{\grayfg}{0.00}\renewcommand{\graybg}{1.00}\else\fi%
  \ifnum #2 > 45 \renewcommand{\grayfg}{0.00}\renewcommand{\graybg}{1.00}\else\fi%
  \ifnum #2 > 50 \renewcommand{\grayfg}{0.00}\renewcommand{\graybg}{1.00}\else\fi%
  \ifnum #2 > 55 \renewcommand{\grayfg}{0.00}\renewcommand{\graybg}{0.82}\else\fi%
  \ifnum #2 > 60 \renewcommand{\grayfg}{0.00}\renewcommand{\graybg}{0.74}\else\fi%
  \ifnum #2 > 65 \renewcommand{\grayfg}{0.00}\renewcommand{\graybg}{0.66}\else\fi%
  \ifnum #2 > 70 \renewcommand{\grayfg}{0.00}\renewcommand{\graybg}{0.58}\else\fi%
  \ifnum #2 > 75 \renewcommand{\grayfg}{1.00}\renewcommand{\graybg}{0.50}\else\fi%
  \ifnum #2 > 80 \renewcommand{\grayfg}{1.00}\renewcommand{\graybg}{0.42}\else\fi%
  \ifnum #2 > 85 \renewcommand{\grayfg}{1.00}\renewcommand{\graybg}{0.34}\else\fi%
  \ifnum #2 > 90 \renewcommand{\grayfg}{1.00}\renewcommand{\graybg}{0.26}\else\fi%
  \ifnum #2 > 95 \renewcommand{\grayfg}{1.00}\renewcommand{\graybg}{0.14}\else\fi%
  \definecolor{cellfg}{gray}{\grayfg}%
  \definecolor{cellbg}{gray}{\graybg}%
  \bscolorcell{white}{cellfg}{cellbg}{2.25em}{1ex}{#1}}

\begin{table*}[tb]
\scriptsize
\centering
\setlength{\tabcolsep}{4.4pt}
\providecommand{\bscolgroupsep}{}
\renewcommand{\bscolgroupsep}{2em}
\caption{Evaluation results of the WSDM Cup 2017 on the test dataset. Performance values are reported in terms of accuracy~(Acc), precision~(P), recall~(R), F-measure, area under the precision-recall curve~(\prauc), and area under curve of the receiver operating characteristic~(\rocauc), as well as with regard to the four data subsets. The darker a cell, the better the performance.}
\label{table-evaluation-results}
\begin{tabular}{@{}lcccccc@{\hspace{\bscolgroupsep}}cc@{\hspace{\bscolgroupsep}}cc@{\hspace{\bscolgroupsep}}cc@{\hspace{\bscolgroupsep}}cc@{}}
\addlinespace
\toprule
\bfseries Approach & \multicolumn{6}{@{}c@{\hspace{\bscolgroupsep}}}{\bfseries Overall performance} & \multicolumn{2}{@{}c@{\hspace{\bscolgroupsep}}}{\bfseries Item head} & \multicolumn{2}{@{}c@{\hspace{\bscolgroupsep}}}{\bfseries Item body} & \multicolumn{2}{@{}c@{\hspace{\bscolgroupsep}}}{\bfseries Registered user} & \multicolumn{2}{@{}c@{}}{\bfseries Unregistered user} \\
\cmidrule(r@{\bscolgroupsep}){2-7}\cmidrule(r@{\bscolgroupsep}){8-9}\cmidrule(r@{\bscolgroupsep}){10-11}\cmidrule(r@{\bscolgroupsep}){12-13}\cmidrule{14-15}
                     & Acc    & P     & R     & F     & \prauc               & \rocauc              & \prauc               & \rocauc              & \prauc               & \rocauc              & \prauc               & \rocauc              & \prauc               & \rocauc              \\
\midrule
META                 & 0.9991 & 0.668 & 0.339 & 0.450 & \bscellA[0.475]{ 47} & \bscellB[0.950]{ 95} & \bscellA[0.648]{ 65} & \bscellB[0.996]{100} & \bscellA[0.387]{ 39} & \bscellB[0.926]{ 93} & \bscellA[0.082]{  8} & \bscellB[0.829]{ 83} & \bscellA[0.627]{ 63} & \bscellB[0.944]{ 94} \\
Buffaloberry         & 0.9991 & 0.682 & 0.264 & 0.380 & \bscellA[0.458]{ 46} & \bscellB[0.947]{ 95} & \bscellA[0.634]{ 63} & \bscellB[0.997]{100} & \bscellA[0.364]{ 36} & \bscellB[0.921]{ 92} & \bscellA[0.053]{  5} & \bscellB[0.820]{ 82} & \bscellA[0.613]{ 61} & \bscellB[0.938]{ 94} \\
Conkerberry          & 0.9990 & 0.675 & 0.099 & 0.173 & \bscellA[0.352]{ 35} & \bscellB[0.937]{ 94} & \bscellA[0.512]{ 51} & \bscellB[0.989]{ 99} & \bscellA[0.281]{ 28} & \bscellB[0.911]{ 91} & \bscellA[0.004]{  0} & \bscellB[0.789]{ 79} & \bscellA[0.538]{ 54} & \bscellB[0.915]{ 91} \\
WDVD (baseline)      & 0.9991 & 0.779 & 0.147 & 0.248 & \bscellA[0.486]{ 49} & \bscellB[0.932]{ 93} & \bscellA[0.668]{ 67} & \bscellB[0.996]{100} & \bscellA[0.388]{ 39} & \bscellB[0.900]{ 90} & \bscellA[0.086]{  9} & \bscellB[0.767]{ 77} & \bscellA[0.641]{ 64} & \bscellB[0.943]{ 94} \\
Honeyberry           & 0.7778 & 0.004 & 0.854 & 0.008 & \bscellA[0.206]{ 21} & \bscellB[0.928]{ 93} & \bscellA[0.364]{ 36} & \bscellB[0.993]{ 99} & \bscellA[0.101]{ 10} & \bscellB[0.893]{ 89} & \bscellA[0.002]{  0} & \bscellB[0.760]{ 76} & \bscellA[0.308]{ 31} & \bscellB[0.819]{ 82} \\
Loganberry           & 0.9285 & 0.011 & 0.767 & 0.022 & \bscellA[0.337]{ 34} & \bscellB[0.920]{ 92} & \bscellA[0.429]{ 43} & \bscellB[0.961]{ 96} & \bscellA[0.289]{ 29} & \bscellB[0.892]{ 89} & \bscellA[0.020]{  2} & \bscellB[0.758]{ 76} & \bscellA[0.487]{ 49} & \bscellB[0.895]{ 89} \\
Riberry              & 0.9950 & 0.103 & 0.483 & 0.170 & \bscellA[0.174]{ 17} & \bscellB[0.894]{ 89} & \bscellA[0.328]{ 33} & \bscellB[0.932]{ 93} & \bscellA[0.113]{ 11} & \bscellB[0.878]{ 88} & \bscellA[0.002]{  0} & \bscellB[0.771]{ 77} & \bscellA[0.378]{ 38} & \bscellB[0.795]{ 79} \\
ORES (baseline)      & 0.9990 & 0.577 & 0.199 & 0.296 & \bscellA[0.347]{ 35} & \bscellB[0.884]{ 88} & \bscellA[0.448]{ 45} & \bscellB[0.973]{ 97} & \bscellA[0.298]{ 30} & \bscellB[0.836]{ 84} & \bscellA[0.026]{  3} & \bscellB[0.627]{ 63} & \bscellA[0.481]{ 48} & \bscellB[0.897]{ 90} \\
FILTER (baseline)    & 0.9990 & 0.664 & 0.073 & 0.131 & \bscellA[0.227]{ 23} & \bscellB[0.869]{ 87} & \bscellA[0.249]{ 25} & \bscellB[0.908]{ 91} & \bscellA[0.182]{ 18} & \bscellB[0.840]{ 84} & \bscellA[0.021]{  2} & \bscellB[0.644]{ 64} & \bscellA[0.387]{ 39} & \bscellB[0.771]{ 77} \\
\bottomrule
\end{tabular}
\end{table*}

\begin{figure*}[t]%
\centering%
\includegraphics[scale=0.51]{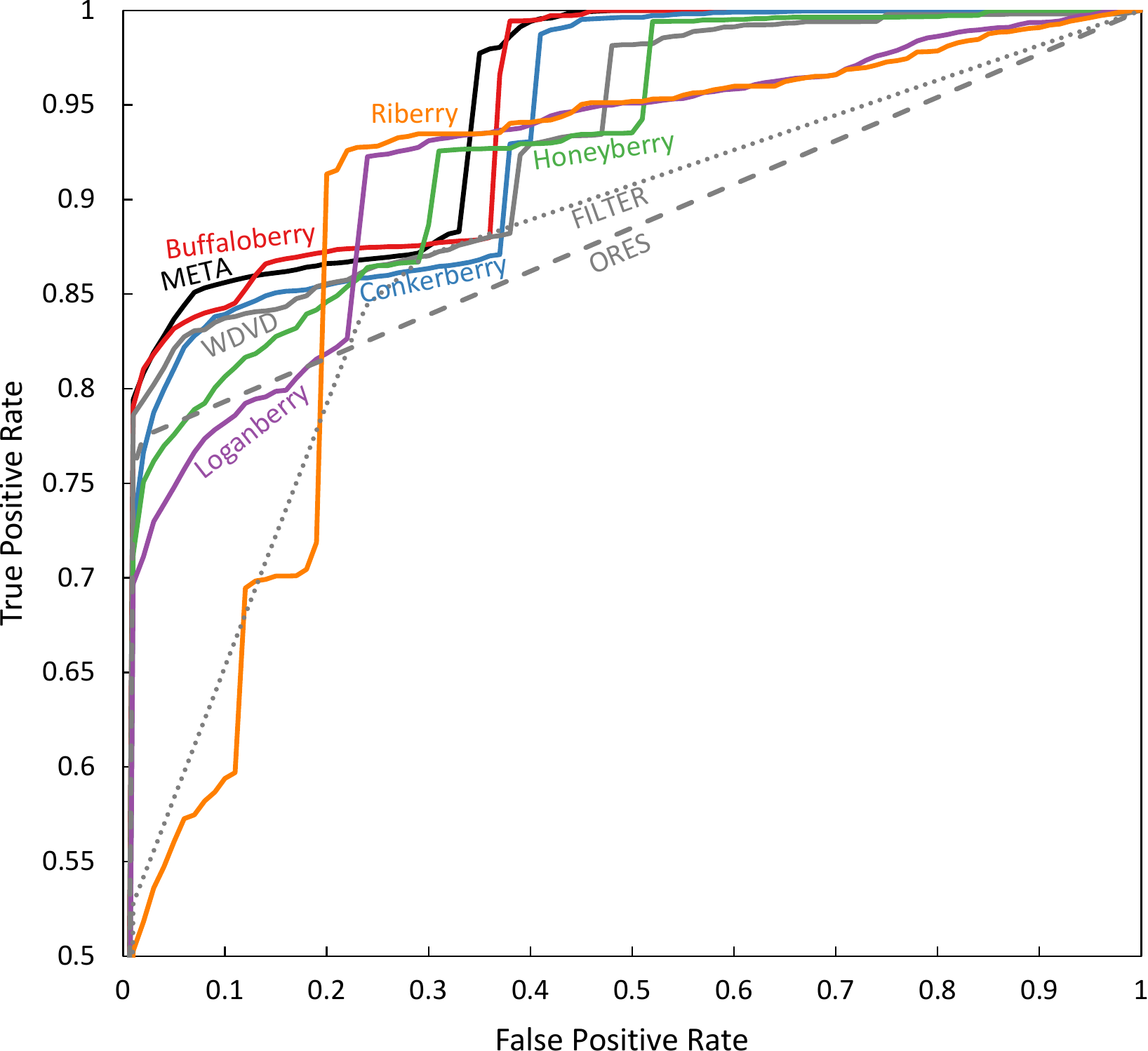}%
\hfill
\includegraphics[scale=0.51]{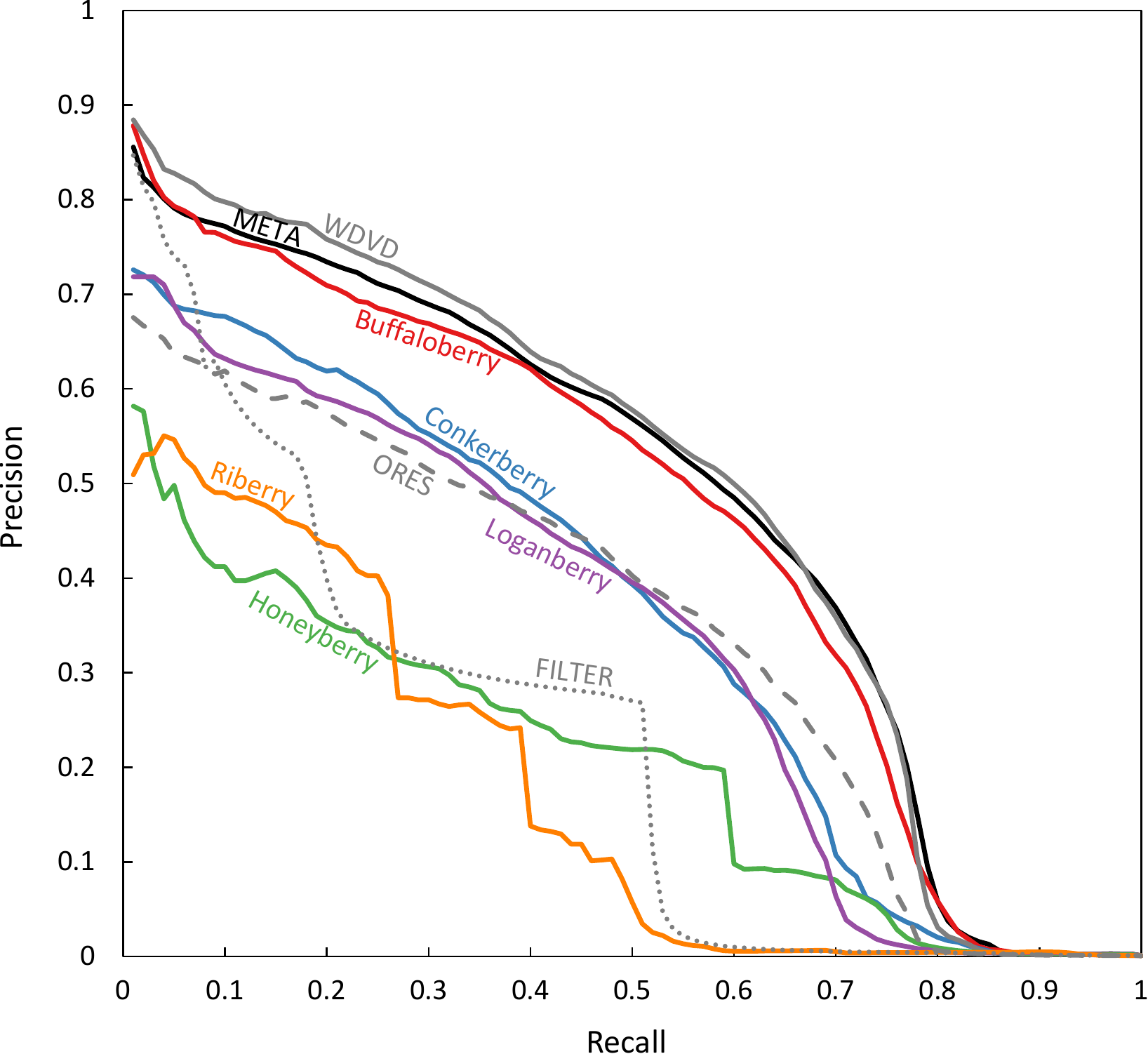}%
\caption{ROC and precision recall curves approaches.}%
\label{curves}%
\end{figure*}

\section{Evaluation}
\label{evaluation}

This section presents an in-depth evaluation of the submitted approaches, including overall performance and performance achieved regarding different data subsets, such as head content vs.\ body content, registered users vs.\ unregistered users, and performance variation over time. Furthermore, we combine the five submitted approaches within an ensemble to get a first idea about the performance that could be achieved if these approaches were integrated.

\subsection{Official Competition Ranking}

The competition phase of the WSDM Cup 2017 officially ended on December~30,~2016, resulting in the following ranking:
\begin{enumerate}
\setlength{\itemsep}{0ex}
\item
Buffaloberry by \citet{Crescenzi2017}
\item
Conkerberry by \citet{Grigorev2017}
\item
Loganberry by \citet{Zhu2017}
\item
Honeyberry by \citet{Yamazaki2017}
\item
Riberry by \citet{Yu2017}
\end{enumerate}

We congratulate the winners! Working versions of these approaches have been successfully deployed within TIRA and evaluated using our aforementioned evaluation framework; three participants also shared their code bases as open source.%
\footnote{\scriptsize https://github.com/wsdm-cup-2017}

To provide a realistic overview of the state of the art at the time of writing, we report the results that were most recently achieved. This is particularly relevant for the authors of the Honeyberry vandalism detector, who found and fixed an error in their approach shortly after the deadline, moving their approach up one rank. Moreover, we include the performances achieved by our own approach as well as that of our two baselines and of the meta approach. The following ranking hence slightly differs from the official one.

\subsection{Evaluation Results}

Table~\ref{table-evaluation-results} and Figure~\ref{curves} give an overview of the evaluation results of the vandalism detection task at WSDM Cup~2017.

\subsubsection{Overall Performance}

The evaluation results shown in Table~\ref{table-evaluation-results} are ordered by \rocauc. The meta approach (\rocauc 0.950) outperforms all other approaches, followed by the winner Buffaloberry (\rocauc 0.947). The least effective approach in terms of \rocauc is the FILTER baseline, achieving 0.869. Note that in our previous work~\citep{Heindorf2016} we reported higher values for \rocauc, which, however, were obtained with a previous version of the dataset. We discuss the apparent differences below.

The \prauc scores are lower and more diverse than the \rocauc scores, which is a consequence of the extreme class imbalance. The WDVD baseline outperforms all approaches in terms of \prauc, including the meta classifier; the ORES baseline outperforms all but two participants. The ranking among the participants changes only slightly: Loganberry and Honeyberry switch places. The main reason for high \prauc scores are features that can signal vandalism with a pretty high precision. For example, WDVD, Buffaloberry, Conkerberry, ORES, and Loganbery are all able to pick up on bad words (\badWordRatio and \bagOfWords) or contain detailed user information (\userFrequency, \userAge, \userVandalismFraction, \userVandalismCount), whereas FILTER and Honeyberry lack the respective features. The performance of Riberry can be explained by the inclusion of features that have previously been found to overfit~\citep{Heindorf2016}.

While \prauc and \rocauc are computed on continuous scores, we also computed accuracy, precision, recall, and F-measure on binary scores at a threshold of~0.5. Honeyberry and Loganberry achieve poorest precision but highest recall. Conkerberry and FILTER achieve poorest recall but high precision. The META approach and Buffaloberry manage the best trade-off in terms of the F-measure. WDVD achieves highest precision at a non-negligible recall. Accuracy correlates with precision due to the high class imbalance. Recall that, since the winner of the competition was determined based on \rocauc, the teams had only little incentive to optimize the other scores. For a real-world applications of classifiers it might be beneficial to calibrate the raw scores to represent more accurate probability estimates and to set the threshold depending on the use case, i.e., adjusting Acc, P, R, and the F-measure.

\subsubsection{Head Content vs.\ Body Content}

Table~\ref{table-evaluation-results} contrasts the approaches' performances on different content types, namely, item heads vs.\ item bodies. Regardless of the metric, all approaches perform significantly better on item heads. We explain this by the fact that vandalism on item heads typically happens at the lexical level (and hence can be detected more easily), e.g., by inserting bad words or wrong capitalization, whereas vandalism on item bodies typically happens at the semantic level, e.g., by inserting wrong facts. In particular, the character and word features focus on textual content as is found in the item head, but there are not many features for structural content. Future work might focus on transferring techniques that are used for the Google knowledge vault~\citep{Dong2014} to Wikidata, such as link prediction techniques to check the correctness of links between items, and Gaussian mixture models to check the correctness of attribute values.

\subsubsection{Registered Users vs.\ Unregistered Users}

Table~\ref{table-evaluation-results} also contrasts the approaches' performances regarding revisions originated from registered users vs.\ revisions from unregistered users. All approaches perform significantly better on revisions from unregistered users, which is in particular reflected by the \prauc scores. Spot-checks suggest that vandalism of unregistered users appears to be rather obvious, whereas vandalism of registered users appears to be more sophisticated and elusive. Moreover, some reverted edits of registered users may also be considered honest mistakes. Telling apart honest mistakes from a (detected) vandalism attempt may be difficult.

\bsfigure[width=\columnwidth]{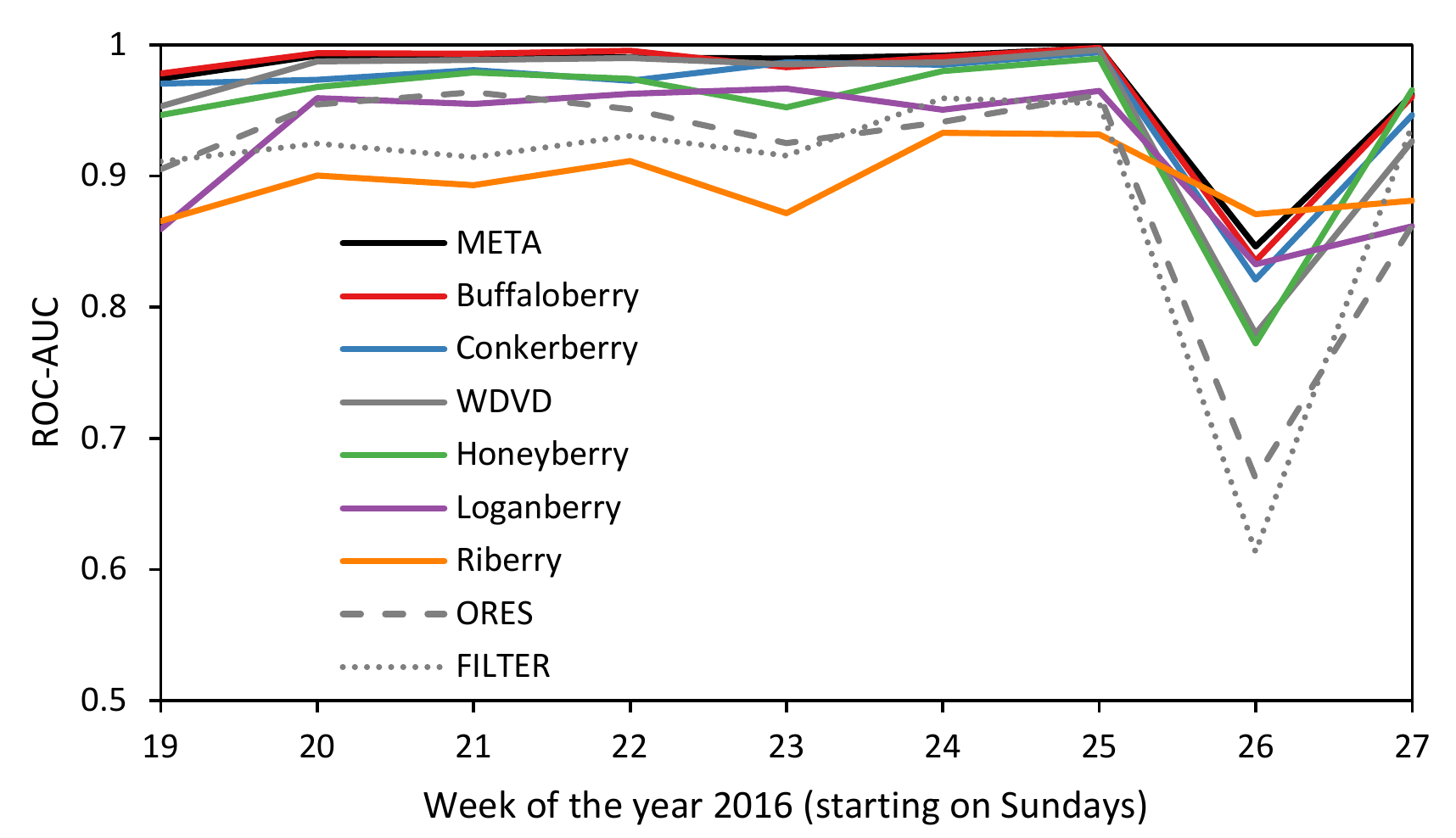}{Performance over time on the test dataset.}

\subsubsection{Performance over Time}

Figure~\ref{plot-roc-over-time} shows the performance of the approaches on the test set over time. Over the first seven weeks (calendar weeks~19 to~25) the approaches' performances remain relatively stable. All approaches were only trained with data obtained up until calendar week~18. Since no drop in the performances is observed, no major changes in the kinds of vandalism seem to have happened in this time frame. However, in calendar week~26 a major performance drop can be observed. The outlier is caused by a single, highly reputable user using some automatic editing tool on June~19, 2016, to create 1,287 slightly incorrect edits (which were rollback-reverted later). Since only 11,043 edits were labeled vandalism in the entire test set of two months, these 1,287~edits within a short time period have a significant impact on the overall performance. 

We were curious to learn about the impact of this data artifact and recomputed the detection performances, leaving out the series of erroneous edits from the user in question. Table~\ref{table-evaluation-results-cleaned} shows the overall performance the approaches would have achieved then: while the absolute performance increases, the ranking among the participants is not affected.

Probably one cannot anticipate such an artifact in the test data, but, with hindsight, we consider it a blessing rather than a curse: it points to the important question of how to deal with such cases in practice. Machine learning-based vandalism detectors become unreliable when the characteristics of the stream of revisions to be classified suddenly changes---errors in both directions, false positives and false negatives, can be the consequence. Ideally, the developers of detectors envision possible exceptional circumstances and provide a kind of exception handling; e.g., flagging a user with suspicious behavior by default for review, regardless whether the respective series of edits is considered damaging or not.

\providecommand{\bscolorcell}[6]{}
\renewcommand{\bscolorcell}[6]{%
  \fcolorbox{#1}{#3}{\parbox{#4}{\centering\vspace{-2pt}\color{#2}\rule[-0pt]{-2pt}{#5}\kern-1pt#6\kern-2.25pt\vspace{-4pt}}}}

\providecommand{\bscellA}[2][]{}
\renewcommand{\bscellA}[2][]{%
  \providecommand{\grayfg}{0.00}%
  \providecommand{\graybg}{1.00}%
  \ifnum #2 = 0  \renewcommand{\grayfg}{0.00}\renewcommand{\graybg}{1.00}\else\fi%
  \ifnum #2 > 0  \renewcommand{\grayfg}{0.00}\renewcommand{\graybg}{0.86}\else\fi%
  \ifnum #2 > 5  \renewcommand{\grayfg}{0.00}\renewcommand{\graybg}{0.82}\else\fi%
  \ifnum #2 > 10 \renewcommand{\grayfg}{0.00}\renewcommand{\graybg}{0.78}\else\fi%
  \ifnum #2 > 15 \renewcommand{\grayfg}{0.00}\renewcommand{\graybg}{0.74}\else\fi%
  \ifnum #2 > 20 \renewcommand{\grayfg}{0.00}\renewcommand{\graybg}{0.70}\else\fi%
  \ifnum #2 > 25 \renewcommand{\grayfg}{0.00}\renewcommand{\graybg}{0.66}\else\fi%
  \ifnum #2 > 30 \renewcommand{\grayfg}{0.00}\renewcommand{\graybg}{0.62}\else\fi%
  \ifnum #2 > 35 \renewcommand{\grayfg}{0.00}\renewcommand{\graybg}{0.58}\else\fi%
  \ifnum #2 > 40 \renewcommand{\grayfg}{0.00}\renewcommand{\graybg}{0.54}\else\fi%
  \ifnum #2 > 45 \renewcommand{\grayfg}{0.00}\renewcommand{\graybg}{0.50}\else\fi%
  \ifnum #2 > 50 \renewcommand{\grayfg}{1.00}\renewcommand{\graybg}{0.46}\else\fi%
  \ifnum #2 > 55 \renewcommand{\grayfg}{1.00}\renewcommand{\graybg}{0.42}\else\fi%
  \ifnum #2 > 60 \renewcommand{\grayfg}{1.00}\renewcommand{\graybg}{0.38}\else\fi%
  \ifnum #2 > 65 \renewcommand{\grayfg}{1.00}\renewcommand{\graybg}{0.34}\else\fi%
  \ifnum #2 > 70 \renewcommand{\grayfg}{1.00}\renewcommand{\graybg}{0.30}\else\fi%
  \ifnum #2 > 75 \renewcommand{\grayfg}{1.00}\renewcommand{\graybg}{0.26}\else\fi%
  \ifnum #2 > 80 \renewcommand{\grayfg}{1.00}\renewcommand{\graybg}{0.22}\else\fi%
  \ifnum #2 > 85 \renewcommand{\grayfg}{1.00}\renewcommand{\graybg}{0.18}\else\fi%
  \ifnum #2 > 90 \renewcommand{\grayfg}{1.00}\renewcommand{\graybg}{0.14}\else\fi%
  \ifnum #2 > 95 \renewcommand{\grayfg}{1.00}\renewcommand{\graybg}{0.10}\else\fi%
  \definecolor{cellfg}{gray}{\grayfg}%
  \definecolor{cellbg}{gray}{\graybg}%
  \bscolorcell{white}{cellfg}{cellbg}{2.25em}{1ex}{#1}}

\providecommand{\bscellB}[2][]{}
\renewcommand{\bscellB}[2][]{%
  \providecommand{\grayfg}{0.00}%
  \providecommand{\graybg}{1.00}%
  \ifnum #2 = 0  \renewcommand{\grayfg}{0.00}\renewcommand{\graybg}{1.00}\else\fi%
  \ifnum #2 > 0  \renewcommand{\grayfg}{0.00}\renewcommand{\graybg}{1.00}\else\fi%
  \ifnum #2 > 5  \renewcommand{\grayfg}{0.00}\renewcommand{\graybg}{1.00}\else\fi%
  \ifnum #2 > 10 \renewcommand{\grayfg}{0.00}\renewcommand{\graybg}{1.00}\else\fi%
  \ifnum #2 > 15 \renewcommand{\grayfg}{0.00}\renewcommand{\graybg}{1.00}\else\fi%
  \ifnum #2 > 20 \renewcommand{\grayfg}{0.00}\renewcommand{\graybg}{1.00}\else\fi%
  \ifnum #2 > 25 \renewcommand{\grayfg}{0.00}\renewcommand{\graybg}{1.00}\else\fi%
  \ifnum #2 > 30 \renewcommand{\grayfg}{0.00}\renewcommand{\graybg}{1.00}\else\fi%
  \ifnum #2 > 35 \renewcommand{\grayfg}{0.00}\renewcommand{\graybg}{1.00}\else\fi%
  \ifnum #2 > 40 \renewcommand{\grayfg}{0.00}\renewcommand{\graybg}{1.00}\else\fi%
  \ifnum #2 > 45 \renewcommand{\grayfg}{0.00}\renewcommand{\graybg}{1.00}\else\fi%
  \ifnum #2 > 50 \renewcommand{\grayfg}{0.00}\renewcommand{\graybg}{1.00}\else\fi%
  \ifnum #2 > 55 \renewcommand{\grayfg}{0.00}\renewcommand{\graybg}{0.82}\else\fi%
  \ifnum #2 > 60 \renewcommand{\grayfg}{0.00}\renewcommand{\graybg}{0.74}\else\fi%
  \ifnum #2 > 65 \renewcommand{\grayfg}{0.00}\renewcommand{\graybg}{0.66}\else\fi%
  \ifnum #2 > 70 \renewcommand{\grayfg}{0.00}\renewcommand{\graybg}{0.58}\else\fi%
  \ifnum #2 > 75 \renewcommand{\grayfg}{1.00}\renewcommand{\graybg}{0.50}\else\fi%
  \ifnum #2 > 80 \renewcommand{\grayfg}{1.00}\renewcommand{\graybg}{0.42}\else\fi%
  \ifnum #2 > 85 \renewcommand{\grayfg}{1.00}\renewcommand{\graybg}{0.34}\else\fi%
  \ifnum #2 > 90 \renewcommand{\grayfg}{1.00}\renewcommand{\graybg}{0.26}\else\fi%
  \ifnum #2 > 95 \renewcommand{\grayfg}{1.00}\renewcommand{\graybg}{0.14}\else\fi%
  \definecolor{cellfg}{gray}{\grayfg}%
  \definecolor{cellbg}{gray}{\graybg}%
  \bscolorcell{white}{cellfg}{cellbg}{2.25em}{1ex}{#1}}

\begin{table}[tb]
\scriptsize
\centering
\setlength{\tabcolsep}{6.2pt}
\providecommand{\bscolgroupsep}{}
\renewcommand{\bscolgroupsep}{2em}
\caption{Evaluation results of the WSDM Cup 2017 on the test dataset without the erroneous edits.}
\label{table-evaluation-results-cleaned}
\begin{tabular}{@{}lcccccc@{}}
\addlinespace
\toprule
\bfseries Approach & \multicolumn{6}{@{}c@{}}{\bfseries Overall performance} \\
\cmidrule{2-7}
                       & Acc    & P     & R     & F     & \prauc               & \rocauc              \\
\midrule                                                                                                %
META              & 0.9992 & 0.668 & 0.384 & 0.487 & \bscellA[0.536]{ 54} & \bscellB[0.988]{ 99} \\
Buffaloberry      & 0.9992 & 0.682 & 0.298 & 0.415 & \bscellA[0.517]{ 52} & \bscellB[0.988]{ 99} \\
WDVD (baseline)   & 0.9992 & 0.779 & 0.167 & 0.274 & \bscellA[0.548]{ 55} & \bscellB[0.980]{ 98} \\
Conkerberry       & 0.9991 & 0.675 & 0.113 & 0.193 & \bscellA[0.398]{ 40} & \bscellB[0.980]{ 98} \\
Honeyberry        & 0.7779 & 0.004 & 0.967 & 0.008 & \bscellA[0.233]{ 23} & \bscellB[0.972]{ 97} \\
Loganberry        & 0.9286 & 0.011 & 0.867 & 0.022 & \bscellA[0.383]{ 38} & \bscellB[0.939]{ 94} \\
FILTER (baseline) & 0.9991 & 0.664 & 0.082 & 0.146 & \bscellA[0.257]{ 26} & \bscellB[0.938]{ 94} \\
ORES (baseline)   & 0.9991 & 0.577 & 0.225 & 0.324 & \bscellA[0.392]{ 39} & \bscellB[0.935]{ 94} \\
Riberry           & 0.9951 & 0.103 & 0.546 & 0.173 & \bscellA[0.196]{ 20} & \bscellB[0.902]{ 90} \\
\bottomrule
\end{tabular}
\end{table}

\section{Reflections on the WSDM Cup}
\label{discussion}

The WSDM Cup~2017 had two tasks for which a total of 140~teams registered, 95~of which ticked the box for participation in the vandalism detection task (multiple selections allowed). This is a rather high number compared with other shared task events. We attribute this success to the facts that the WSDM conference is an A-ranked conference, giving the WSDM Cup a high visibility, that the vandalism detection task was featured on Slashdot,%
\footnote{https://developers.slashdot.org/story/16/09/10/1811237}
and that we attracted sponsorship from Adobe, which allowed us to award cash prizes to the three winning participants of each task. However, only 35~registered participants actually engaged when being asked for their operating system preferences for their virtual machine on~TIRA, 14~of which managed to produce at least one run, whereas the remainder never used their assigned virtual machines at all. In the end, five teams made a successful submission by running their software without errors on the test dataset.

Why did so many participants decided to drop out on this task? We believe that the comparably huge size of the dataset as well as difficulties in setting up their approach in our evaluation platform are part of the reason: each approach had to process gigabytes of data by implementing a client-server architecture, and all of that had to be deployed on a remote virtual machine. The requirement to submit working software, however, may not have been the very main cause since the retention rate of our companion task was much higher. Rather, the combination of dataset size, real-time client-server processing environment, and remote deployment is a likely cause. Note that the vandalism detection task itself demanded for this scale of operations, since otherwise it would have been easy to cheat, which is particularly a problem when cash prizes are involved. Finally, the provided baseline systems were already competitive, so that the failure to improve upon them may have caused additional dropouts.

The WSDM Cup taught us an important lesson about the opportunities and limitations of shared tasks in general and about evaluation platforms and rule enforcement in particular. On the one hand, competitions like ours are important to rally followers for a given task and to create standardized benchmarks. On the other hand, shared tasks are constrained to a relatively short period of time and create a competitive environment between teams. I.e., it becomes important to implement a good trade-off in the evaluation setup in order to prevent potential cheating and data leaks, while, at the same time, placing low hurdles on the submission procedure. A means towards this end might be standardized evaluation platforms that are widely used for a large number of shared tasks. While there are already platforms like TIRA or Kaggle, we are not aware of a widely used evaluation platform for time series data, serving teams with one test example after the other and providing a sandboxing mechanism of the submitted softwares to prevent data leaks. Moreover, there definitely is a trade-off between enforcing strict rules on the one side and scientific progress on the other. For example, only two teams had made a successful submission by the original deadline, while other teams were still struggling with running their approaches. In this case, we erred on the side of scientific progress by additional submissions in favor of overly strict rules and gave all teams a short deadline extension of 8 days, accepting some discussion about the deadline's extensions fairness.

\section{Conclusion and Outlook}

This paper gives an overview of the five vandalism detection approaches submitted to the WSDM Cup~2017. The approaches were evaluated on the new Wikidata Vandalism Corpus~2016, which has been specifically compiled for the competition. Under a semi-automatic detection scenario, where newly arriving revisions are ranked for manual review, the winning approach from Buffaloberry \citet{Crescenzi2017} performs best. Under a fully automatic detection scenario, where the decision whether or not to revert a given revision is left with the classifier, the baseline approach WDVD by \citet{Heindorf2016} still performs best. Combining all approaches within a meta classifier yields a small improvement; however, the feature set seems to be the performance-limiting factor.

All approaches build upon the WDVD baseline, proposing only few additional features. I.e., for the future it is interesting to develop and explore fundamentally different feature sets. E.g., building upon the work on knowledge graphs, technology for link prediction and value range prediction should be investigated. Building upon the work of other user-generated content, also psychologically motivated features capturing a user's personality and state of mind appear promising.

\section*{Acknowledgments}

We thank Adobe for sponsoring the WSDM Cup, and Wikimedia Germany for supporting our task. Our special thanks go to all participants for their devoted work.

{
\fontsize{8.5pt}{9.5pt}\selectfont
\raggedright

}

\balancecolumns

\end{document}